# An exact solution of the lubrication equations for the Oldroyd-B model in a hyperbolic channel[1]

**Kostas D. Housiadas**

*Department of Mathematics, University of the Aegean, Karlovassi, Samos, 83200, Greece*

**Abstract**

An exact similarity solution of the lubrication equations for the steady isothermal, incompressible flow of a viscoelastic Oldroyd-B fluid in a contracting and symmetric hyperbolic channel is derived. The solution is valid for finite (of order unity) values of the Deborah number, *De* (the ratio of the polymer's longest relaxation time to a characteristic residence time of the fluid in the channel), all values of the polymer viscosity ratio, *η* (the ratio of polymer viscosity to the total viscosity of the fluid), as well as to typical values of the contraction ratio, Λ (the ratio of the channel height at the inlet to the channel height at the outlet). The solution, the significance of which for the hyperbolic geometry is analogous to the classic Poiseuille solution in a straight channel, precisely satisfies the exact analytical solution of the Oldroyd-B model along the walls of the channel, but does not satisfy any initial conditions for the polymer extra-stresses at the inlet of the hyperbolic section of the channel. The new exact solution is given in terms of the streamfunction and is used in the momentum balance to derive a non-linear ordinary differential equation with an unknown function which corresponds to a modified fluid velocity. The final equation is solved numerically using a fully spectral method with a Galerkin-type approach to calculate the unknown function and the pressure gradient. The exact solution for the polymer extra-stresses allows for deriving a variety of expressions for the average pressure-drop in the channel. The most accurate expression is that resulting from the mechanical energy decomposition of the flow. The range of validity of the approximate analytical formula for the average pressure drop is determined in terms of *De*, Λ and *η,* and confirmed by numerical pseudospectral simulations of the full lubrication equations. In all cases, a decrease in the average pressure drop compared to the Newtonian value with increasing *De* and/or *η* is predicted.

---

## 1. Introduction

Pressure-driven flows of viscoelastic fluids in narrow and long tubes (planar channel or axisymmetric circular pipes) are widely encountered in industrial processes, such as extrusion (Pearson 1985; Tadmor & Gogos 2013), in applications such as microfluidic extensional rheometers (Ober *et al.* 2013), in devices for subcutaneous drug administration (Allmendinger *et al.* 2014; Fischer *et al.* 2015) and many others. More specifically, contraction flows through hyperbolically-shaped tubes have attracted a lot of attention because of the almost linear velocity profile along the main flow direction, developed on the midplane for planar geometries or along the axis of symmetry for axisymmetric ones, which leads to an almost constant extensional rate. These flows have often been used to measure the extensional viscosity of the fluid by relating the average pressure drop along the tube to the flow rate (James, Chandler & Armour 1990; Collier *et al.* 1998; Feigl *et al.* 2003; Koppol *et al.* 2009; Wang & James 2011; Ober *et al.* 2013; Nyström *et al.* 2016; Kim *et al.* 2018). Note, however, that these flows are primarily shear-dominated almost everywhere at each cross-section, except at the midplane for the planar geometry and along the axis of symmetry for the 3D axisymmetric geometry, where they become purely extensional. (Housiadas & Beris 2024a,b,c).

Most of the available experimental data in literature for viscoelastic fluids in contraction flows show an increase of the pressure drop with increasing the fluid viscoelasticity. These data mainly concern with flows at high Weissenberg number, or in geometric configurations with singular points such as the sudden contraction-expansion planar 2-D channels (Nigen & Walters 2002), 3-D square channels (Sousa *et al.* 2009), or the contraction-expansion circular axisymmetric tube (Rothstein & McKinley 1999, 2001; Nigen & Walters 2002). In these cases, secondary motion, vortices, and/or elastic instabilities arise which cause the increase of the pressure drop in the tube. These flows and geometries are different from the pure hyperbolic case studied here, in which no corners exist, and the fluid enters the hyperbolic section of the channel smoothly in a fully developed state without vortical or secondary motion.

The theoretical study of viscoelastic flows is highly complex, primarily due to the nonlinear nature of the appropriate governing equations. This complexity persists even when using basic constitutive models, such as the Upper Convected Maxwell (UCM) and Oldroyd-B models, and under simplified flow conditions (isothermal, steady, laminar, and creeping flows). Another challenge is the ability of constitutive models to accurately capture the macromolecular response to flow deformation across a wide range of flow conditions, geometries, and fluid properties. Finally, it is now well accepted that under certain conditions, viscoelastic instabilities arise, which in some cases may even lead to turbulence (Shaqhef 1996; Datta *et al.* 2022). These instabilities typically appear under high levels of fluid viscoelasticity and are not exclusively associated with curved streamlines or closed geometries, e.g., periodic (Shaqfeh 1996).

Thus, the development and use of theories or techniques that can simplify the original governing equations without overlooking major features of the flow is of crucial importance. Such a theory is the lubrication theory used widely for the modeling of thin fluid films (Szeri 2005; Langlois & Deville 2014; Ockendon & Ockendon 1995; Leal 2007), the motion of particles near surfaces (Goldman *et al.* 1967; Stone 2005), the flow in microchannels with known geometry (Stone *et al.* 2004; Plourabouté *et al.* 2004; Amyot *et al.* 2007; Tavakol *et al.* 2017), and generally for the investigation of slow flows in narrow and confined geometries.

For instance, in inertialess Newtonian flow within confined tubes (planar channels or axisymmetric pipes), the classic lubrication theory significantly simplifies the momentum balance by assuming that the fluid motion in the vertical direction is much slower than the main flow direction and by neglecting derivatives of the velocity components with respect to the main flow direction. Consequently, boundary conditions for the velocity field are neither needed nor can they be imposed at the tube's inlet and outlet, thus eliminating entrance and exit effects. These simplifications, which are justified because the aspect ratio of the tube is small, reveal that the pressure gradient along the main flow direction is independent of the transverse (or radial, or wall-normal) direction. Physically, this implies that in long tubes, entrance and exit effects are of minor significance and can be disregarded as a first approximation. Furthermore, by applying a total force balance on the system for inertialess Newtonian flow in the absence of external forces or torques, it can be shown that the pressure drop required to maintain a constant flow rate arises solely from the viscous forces exerted by the fluid on the tube wall(s). Equivalently, with the aid of the mechanical energy balance on the flow system, the required pressure drop can be shown to result from energy losses in the flow due to pure viscous dissipation.

For a viscoelastic fluid, however, the corresponding pressure drop resulting from the total force balance on the system is attributed not only to viscous forces on the tube wall(s) but also to an additional contribution from the fluid's viscoelasticity, as demonstrated by Housiadas & Beris (2023; 2024a,c) and, independently, by Hinch, Boyko & Stone (2024) and Boyko, Hinch & Stone (2024). Note however, that these authors used slightly different approaches: Housiadas & Beris (2023; 2024a,c,d) utilized the exact analytical solution of the Oldroyd-B model at the walls, the integral constraint due to fluid's incompressibility, and the integrated momentum balance along both spatial directions, while Hinch, Boyko & Stone (2024) and Boyko, Hinch & Stone (2024) used the integral constraint due to fluid's incompressibility, which integrated by parts twice, and use of the momentum balance without accounting for the exact solution of the Oldroyd-B model at the walls. Similarly, through a mechanical energy balance of the flow, the pressure drop can be decomposed into a component due to viscous dissipation and another due to elastic dissipation, alongside with the work done by elastic forces (Housiadas & Beris 2024c; Hinch, Boyko & Stone 2024, Sialmas & Housiadas 2025).

Application of the lubrication theory to study viscoelastic lubricant flows in tubes made by solid walls has been performed in the area of tribology by Tichy (1996), who was the first to derive the lubrication equations for the UCM model, followed by others such as Zhang, Matar & Craster (2002), Li (2014), Gamaniel *et al.* (2021), Ahmed & Biancofiore (2021, 2023), and Sari *et al.* (2024). Additionally, theoretical investigation of the effect of viscoelasticity, augmented by numerical simulations, on contracting channels (mainly with hyperbolic geometry) was carried out by Perez-Salas *et al.* (2019), Boyko & Stone (2022), Housiadas & Beris (2023, 2024a,b,c,d), Hinch *et al.* (2024), Boyko *et al.* (2024). Even, however, under the lubrication approximation, the final governing equations remain non-linear, and thus exact analytical solutions for flows in non-uniform geometries are not available in the literature. As is well-known in the non-Newtonian fluid mechanics community, theoretical analytical results can be obtained only for weakly viscoelastic fluids by applying regular perturbation expansion(s) for all the field variables in terms of the Weissenberg or Deborah numbers. Therefore, exact analytical solutions of the constitutive equation(s) used to model the response of the polymer molecules to the flow deformation are invaluable because they advance substantially the theoretical framework of obtaining accurate results for the major quantities of theoretical, practical or engineering interest.

The purpose of this work is to investigate analytically the steady creeping flow of a viscoelastic fluid in a hyperbolic symmetric channel by deriving an exact solution for the polymer extra-stress tensor within the framework of the classic lubrication limit, valid in the limit of a vanishingly small aspect ratio of the channel. The Oldroyd-B model is utilized only, since is the most fundamental non-linear differential constitutive model in viscoelasticity, and significant analytical progress can be achieved. Additionally, previous works in the literature for the planar hyperbolic geometry (Housiadas & Beris 2023, 2024a,b,c) using more realistic and more non-linear macroscopic constitutive models such as the Giesekus, Phan-Thien & Tanner and FENE-P models, showed only minor differences with the corresponding UCM/Oldroyd-B results for small or intermediate levels of fluid's elasticity. This is attributed to the lubrication approximation within which most of the non-linear terms of the polymer extra-stress tensor in the constitutive equations are eliminated (Housiadas & Beris 2023; 2024a,b,c).

The rest of the paper is organized as follows. In section 2, the definition of the flow problem is presented, and dimensionless variables based on the lubrication theory are introduced. The final equations at the classic lubrication limit along with the accompanying auxiliary conditions are also presented. In Section 3, the lubrication equations are formulated in terms of the streamfunction, new modified components of the polymer extra-stress tensor, and mapped spatial coordinates, leading to a new set of partial differential equations. In Section 4, the equation of the pressure gradient at each cross of the channel is derived, along with the equation for the average pressure drop required to maintain a constant flowrate thought the channel. In section 5, the new set of lubrication equations is solved using three methods. The first method is a standard perturbation scheme valid for weakly viscoelastic fluids, while the second method is a high-accuracy pseudospectral numerical method. The third method assumes a specific dependence of the field variables on the new mapped coordinates allowing for the derivation of an exact solution of the Oldroyd-B in terms of the streamfunction. Section 6 focuses on the average pressure drop derived from the new exact solution, with its accuracy assessed against the corresponding high-order perturbation and numerical solutions of the new set of lubrication equations. A variety of analytical formulas for the average pressure drop are also derived. Section 7 provides a discussion of the properties of the new solution, its limiting cases, its range of applicability, and the values of the dimensionless parameters that appear in the lubrication equations. Finally, the main conclusions and suggestions for future research problems using the new solution are presented in Section 8.

## 2. Problem formulation

In Figure 1, a symmetric channel with respect to the midplane is depicted. It consists of three segments; an entrance region with constant height $2 h_0^*$, a varying region with length $\ell^*$, height $2 h_0^*$ at the inlet and height $2 h_f^*$ at the outlet, and an exit region with constant height $2 h_f^*$; throughout the paper a star superscript denotes dimensional quantity. We also assume a constant volumetric flow rate, $Q^*$, and that the flow in the entrance region of the channel is steady and fully developed. Moreover, creeping flow conditions are considered, i.e. the flow is always slow, and the transition from the entrance region to the hyperbolic section is smooth to prevent flow instabilities at the inlet (such as secondary vorticial motion); Binding 1988; Nigen & Walters 2002; Sousa *et al.* 2009; Rothstein & McKinley 1999, 2001; Lubansky *et al.* 2007; Lopez-Aguilar *et al.* 2016). This is very different from, for instance, the inlet conditions of the sudden contraction-expansion geometry studied by Rothstein & McKinley (1999, 2001), or the abrupt axisymmetric contraction studied by Lubansky *et al.* 2007, in which major vortices appear in the corner region right before the contraction section of the tube.

First, we define the aspect ratio of the varying region of the channel, the ratio of the inlet to the outlet heights, and half the average inlet velocity, respectively:

$$\varepsilon \equiv \frac{h_0^*}{\ell^*}, \quad \Lambda \equiv \frac{h_0^*}{h_f^*}, \quad u_c^* \equiv \frac{Q^*}{h_0^*}. \tag{2.1}$$

For a narrow channel such as that shown in Figure 1 the aspect ratio is small, $\varepsilon < 1$, a feature that is exploited for the derivation of the lubrication equations at the limit of a vanishingly small aspect ratio. For $\Lambda > 1$, a contracting channel is described, while for $0 < \Lambda < 1$ an expanding one; in the former case Λ will be referred to as the contraction ratio. In the experiments, the contraction ratio is usually in the range 4-10 for the 2D planar geometry, and is even lower, between 3 and 7 for the 3D axisymmetric geometry (James & Roos 2021; James & Tripathi 2023). This limitation arises because the very large stresses developed near the channel exit can cause channel fracture (see Section 7 for more details).

We consider a complex incompressible viscoelastic fluid (a polymeric material dissolved into a Newtonian solvent, i.e. a dilute polymer solution) with longest relaxation time $\lambda^*$, and we recognize two distinct time scales associated with the flow; an inverse shear-rate $h_0^*/u_c^*$, and a characteristic residence time of the fluid in the channel $\ell^*/u_c^*$. Based on these characteristic times, we define the Weissenberg and Deborah numbers, as well as the modified Deborah number (Housiadas & Beris 2023; Housiadas & Beris 2024a,b,c):

$$Wi \equiv \frac{\lambda^*}{h_0^*/u_c^*} = \frac{\lambda^* u_c^*}{h_0^*}, \quad De \equiv \frac{\lambda^*}{\ell^*/u_c^*} = \frac{\lambda^* u_c^*}{\ell^*}, \quad De_m \equiv \frac{\lambda^* u_c^*}{\ell^*}\left(\frac{h_0^*}{h_f^*}-1\right), \quad (2.2)$$

from where we observe that $De = \varepsilon\, Wi$ and $De_m = \varepsilon\, Wi(\Lambda - 1)$; the latter is a dimensionless group that combines both viscoelastic and geometrical characteristics of the flow and is especially useful for interpreting the results. Moreover, it has been revealed that the hyperbolic geometry can be used experimentally to evaluate the steady elongational viscosity of the fluid only in the range $0 < 3De_m/4 < 1/2$ for the 2D planar geometry (Housiadas & Beris 2024b,c), i.e., for $0 < De_m < 2/3$, with similar conclusions having been made for the 3D axisymmetric case. Since the analysis performed here assumes a confined and narrow geometry, $0 < \varepsilon < 1$, and in order to be able to observe the effect of viscoelasticity, we assume that $De$ is finite when $\varepsilon$ becomes very small, which in turn implies that $Wi$ is large, i.e. $De = O(1)$ and $Wi = O(1/\varepsilon)$ for $\varepsilon \ll 1$. Note, however, that for steady, inertialess flow in the channel, the residence time of the fluid elements increases with distance from the midplane, approaching infinity at the walls due to the no-slip condition. Therefore, from a Lagrangian perspective, fluid elements entering the channel experience different dynamics and evolution as they move toward the channel exit; the closer they are to the walls, the lower the effective Deborah number. This observation also demonstrates that very large values of the Deborah number are inconsistent with the correct underlying physics for this type of flow, and a high Deborah number analysis, along with the corresponding results, should be considered with extreme caution.

Notice that the definitions of the Deborah and Weissenberg numbers given in Eq. (2.2) are similar, namely these two dimensionless groups are related through a geometric factor; *De* is based on the channel's length, while *Wi* is based on its half-height, and therefore Despite their similarity, they have different physical meaning and quantify different effects (Pipkin 1986; Macosco 1994; Poole 2012; Dealy 2012; Salvatore *et al.* 2024); *Wi* measures the ratio of the fluid's relaxation time to a characteristic time of the process related to the reciprocal shear-rate, while *De* measures the ratio of the fluid's relaxation time to the fluid's characteristic residence time in the channel. In other words, *De* shows the relative elastic to viscous response of the material. High *Wi* values indicate strong viscoelasticity, while high *De* values correspond to materials deforming practically like elastic solids. Note that in many flow problems where only a limited number of characteristic scales are available, these dimensionless numbers may coincide. An important consequence of flows for large values of both *Wi* and *De* numbers is that the well-known differential models such as the UCM/Oldroyd, Giesekus, FENE-P (and its modification, FENE-CR), and Phan-Thien & Tanner models are not valid.

It is also well known that large values of both the Weissenberg and Deborah numbers can lead to instabilities, which may ultimately result in viscoelastic turbulence (Shaqfeh, 1996; Datta *et al.*, 2022). Under such conditions, simplified mathematical models, such as the lubrication equations used here, are insufficient to accurately describe the flow dynamics. This situation, however, is common in fluid mechanics. For instance, in steady Poiseuille flow, whether in a 2D planar channel or a 3D axisymmetric pipe, increasing the Reynolds number leads to the onset of instabilities and eventually turbulence, rendering the classical Poiseuille predictions invalid.

For all these reasons, for the steady creeping lubrication flow in a hyperbolic channel studied here, the analysis is restricted to low to intermediate values of the Deborah number. Notice that in the experiments conducted by James & Roos (2021) in a hyperbolic 3D axisymmetric geometry, the maximum attainable value for the Deborah number was 4.6 (see Figure 9, in their paper). Accounting for the differences in the characteristic scales this corresponds to *De*=0.145 (see the discussion in subsection 7.3. for more details). It is also worth noting that some authors have employed inverse definitions for these dimensionless numbers. For example, Rothstein and McKinley (1999) report Deborah numbers equal or larger than six for an axisymmetric sudden contraction/expansion geometry; the latter differs a lot from the hyperbolic geometry. However, their Deborah number is based on the tube diameter, not its length, meaning that their reported Deborah number aligns with the Weissenberg number as defined in the present study.

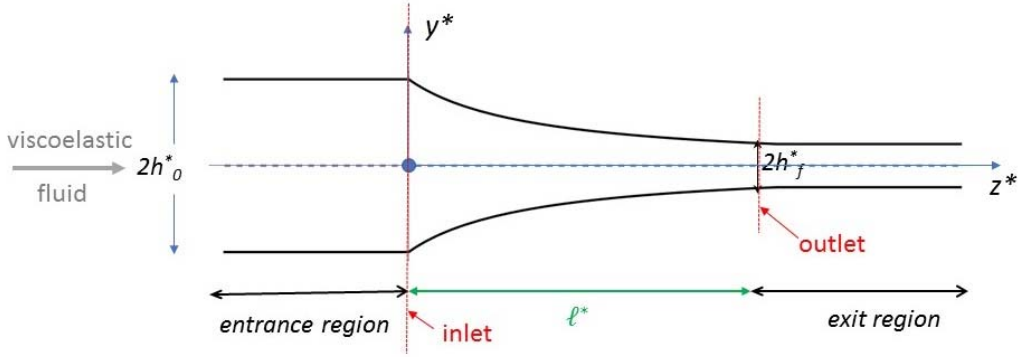

**Figure 1:** Geometry and Cartesian coordinate system ($y^*, z^*$) for a symmetric hyperbolic channel

### *2.1. Governing equations*

We consider the isothermal and steady flow of a viscoelastic fluid which consists of a Newtonian solvent with constant shear viscosity $\eta_s^*$ and a polymeric material with constant shear viscosity $\eta_p^*$. We use a Cartesian coordinate system $(z^*, y^*, x^*)$ to describe the flow field, where $z^*$ is the main flow direction, $y^*$ is the vertical direction, and $x^*$ is the neutral direction; $\mathbf{e}_x, \mathbf{e}_z$ and $\mathbf{e}_y$ are the unit vectors in the $x^*$, $z^*$, and $y^*$ directions, respectively. The origin of the coordinate system is placed on the center of the inlet cross-section with radius $h_0^*$ (see Figure 1). The walls of the hyperbolic channel are described by the shape function $H^* = H^*(z^*) > 0$ for $0 \leq z^* \leq \ell^*$, i.e., $y^* = \pm H^*(z^*)$, where

$$H^*(z^*) = \frac{h_0^*}{1 + (h_0^*/h_f^* - 1)(z^*/\ell^*)}. \tag{2.3}$$

For $h_0^* = h_f^*$ one gets a straight channel, for $h_0^* > h_f^*$ a contracting channel, and for $h_0^* < h_f^*$ an expanding channel. Note that here we are interested in the contracting case only. The reason for choosing the hyperbolically-shaped channel, as described by Eqs. (2.3) is because it leads to a fully symmetric flow field and simultaneously generates an almost constant elongation rate along the midplane of the channel. Consequently, the velocity field at the midplane/axis of symmetry is linear, i.e. this specific form of the shape function helps to the formation of a flow field which resembles that for pure homogenous planar 2D elongation [Tanner 2000; Housiadas & Beris 2024b] or constant uniaxial 3D extension [Tanner 2000; Housiadas & Beris 2024b,d]. Because of these flow characteristics the hyperbolic channels (and pipes too) have been used frequently for the experimental determination of the extensional viscosity of complex fluids [Coswell 1972, 1978; Jones *et al.* 1987; Binding & Walters 1988; Binding & Jones 1989; Aboubacar *et al.* 2002; Alves *et al.* 2003; Oliveira *et al.* 2007; Kamerkar & Edwards 2007; Aguayo *et al.* 2008; Campo-Deano *et al.* 2011; Ober *et al.* 2013; Keshavarz & McKinley 2016; Lee & Muller 2017].

The velocity vector in the flow domain is denoted by $\mathbf{u}^* = V^*(y^*,z^*)\mathbf{e}_y + U^*(y^*,z^*)\mathbf{e}_z$ and the total pressure by $P^* = P^*(y^*,z^*)$. Using the unit tensor $\mathbf{I}$, the rate of deformation tensor $\dot{\boldsymbol{\gamma}}^* = \nabla^*\mathbf{u}^* + (\nabla^*\mathbf{u}^*)^T$ where $\nabla^*$ is the gradient operator, and the viscoelastic extra-stress tensor $\boldsymbol{\tau}^*$ we define the total momentum tensor:

$$\mathbf{T}^* := -P^*\mathbf{I} + \eta_s^* \dot{\boldsymbol{\gamma}}^* + \boldsymbol{\tau}^*. \qquad (2.4)$$

In the absence of any external forces and torques, the conservation equations that govern the flow in the channel are the mass and momentum balances, respectively:

$$\nabla^* \cdot \mathbf{u}^* = 0, \quad \nabla^* \cdot \mathbf{T}^* = \mathbf{0}. \qquad (2.5)$$

The response of the polymeric material to the flow deformation is modeled using the fundamental non-linear differential Oldroyd-B model:

$$\boldsymbol{\tau}^* + \lambda^*\left(\mathbf{u}^* \cdot \nabla^*\boldsymbol{\tau}^* - \boldsymbol{\tau}^* \cdot \nabla^*\mathbf{v}^* - (\nabla^*\mathbf{v}^*)^T \cdot \boldsymbol{\tau}^*\right) = \eta_p^* \dot{\boldsymbol{\gamma}}^*, \qquad (2.6)$$

Eq. (2.6) can also be formulated equivalently in terms of the dimensionless, symmetric and second-order conformation tensor $\mathbf{C}$ (Bird, Armstrong & Hassager 1987; Beris & Edwards 1996). For the Oldroyd-B model, $\mathbf{C}$ is linearly related to $\boldsymbol{\tau}^*$ as $\boldsymbol{\tau}^* = \eta_p^*(\mathbf{C}-\mathbf{I})/\lambda^*$, where $\mathbf{I}$ is the unit tensor.

The domain of definition of Eqs. (2.5)-(2.6) is:

$$\Omega^* = \left\{(z^*, y^*) \mid, 0 < z^* < \ell^*, -H^*(z^*) < y^* < H^*(z^*)\right\}.$$

The governing equations are solved with no-slip and no-penetration boundary conditions along the wall of the channel:

$$U^*(H^*(z^*), z^*) = V^*(H^*(z^*), z^*) = 0 \quad \text{at} \quad 0 \le z^* \le \ell^*. \qquad (2.7a)$$

Symmetry conditions at the midplane can also be satisfied at $y^* = 0$:

$$V^*(0, z^*) = \left.\frac{\partial U^*}{\partial y^*}\right|_{y^*=0} = \tau_{yz}^*(0, z^*) = 0 \quad \text{at} \quad 0 \le z^* \le \ell^*, \qquad (2.7b)$$

The integral constraint of mass at any distance from the inlet is also utilized:

$$Q^* = \int_{-H^*(z^*)}^{H^*(z^*)} U^*(y^*, z^*) dy^* = \text{constant} \text{ at } 0 \leq z^* \leq \ell^*. \tag{2.7c}$$

Finally, a datum pressure, $p_{ref}^*$, is chosen at the wall of the outlet cross-section:

$$p_{ref}^* = P^*(H^*(\ell^*), \ell^*). \tag{2.7d}$$

Note that when the Newtonian solvent is absent ($\eta_s^* = 0$) the governing equations reduce to the Upper Convected Maxwell (UCM) model, or to the Oldroyd-B model when both the Newtonian solvent and the polymer molecules are present, $\eta_s^*, \eta_p^* > 0$.

## 2.2. Lubrication approximation

Dimensionless variables are introduced based on the lubrication theory by scaling $z^*$ by $\ell^*$, $y^*$ and $H^*$ by $h_0^*$, $U^*$ by $u_c^*$, and $V^*$ by $\varepsilon u_c^*$. The pressure difference $P^* - P_{ref}^*$ is scaled by $(\eta_s^* + \eta_p^*) u_c^* \ell^* / h_0^{*2}$ (Tavakol *et al.* 2017, Boyko & Stone 2022; Housiadas & Beris 2023, 2024a,b,c,d; Boyko *et al.* 2024; Hinch *et al.* 2024). For the viscoelastic extra-stress components, $\tau_{zz}^*, \tau_{yz}^*$, and $\tau_{yy}^*$ the characteristic scales are $\eta_p^* u_c^* \ell^* / h_0^{*2}$, $\eta_p^* u_c^* / h_0^*$, and $\eta_p^* u_c^* / \ell^*$, respectively. Using the scales and dimensionless coordinates and variables, one derives the complete form of governing equations which can be found in the literature (Boyko & Stone, 2022; Housiadas & Beris 2023; Housiadas & Beris 2024a-d; Boyko *e al.* 2024; Hinch *et al.* 2024). In these equations, the aspect ratio of the channel appears in even powers only. For a long and narrow channel, the aspect ratio is small which implies that from an asymptotic point of view, all terms in the dimensionless governing equations multiplied to $\varepsilon^2$ or $\varepsilon^4$ are much smaller compared to the other terms and can be ignored as a first approximation, leading to the lubrication equations for an Oldroyd-B model:

$$\frac{\partial U}{\partial z} + \frac{\partial V}{\partial y} = 0, \tag{2.8}$$

$$0 = -\frac{\partial P}{\partial z} + (1-\eta)\frac{\partial^2 U}{\partial y^2} + \eta\left(\frac{\partial \tau_{zz}}{\partial z} + \frac{\partial \tau_{yz}}{\partial y}\right), \tag{2.9}$$

$$0 = -\frac{\partial P}{\partial y}, \tag{2.10}$$

$$\tau_{zz} + De\left(\frac{D\tau_{zz}}{Dt} - 2\tau_{zz}\frac{\partial U}{\partial z} - 2\tau_{yz}\frac{\partial U}{\partial y}\right) = 0, \tag{2.11}$$

$$\tau_{yz} + De\left(\frac{D\tau_{yz}}{Dt} - \tau_{zz}\frac{\partial V}{\partial z} - \tau_{yy}\frac{\partial U}{\partial y}\right) = \frac{\partial U}{\partial y}, \qquad (2.12)$$

$$\tau_{yy} + De\left(\frac{D\tau_{yy}}{Dt} - 2\tau_{yz}\frac{\partial V}{\partial z} - 2\tau_{yy}\frac{\partial V}{\partial y}\right) = 2\frac{\partial V}{\partial y}, \qquad (2.13)$$

where $D/Dt \equiv U(\partial/\partial z) + V(\partial/\partial y)$ is the convective derivative at steady state. The same type of lubrication analysis has been performed before for the 2D planar case by Boyko & Stone (2022), Ahmed & Biancofiore (2021; 2023), Housiadas & Beris (2023; 2024a,b,c), and Hinch, Boyko & Stone (2024), and for the 3D axisymmetric case in a pipe by Housiadas & Beris (2024d) and Sialmas & Housiadas (2025). Eqs (2.8)-(2.13) can also be formulated equivalently in terms of the components of the conformation tensor, as previously employed by Boyko & Stone (2022), Hinch, Boyko & Stone (2024), and discussed by Housiadas & Beris (2023; 2024d) considering the expressions $\tau_{zz} = c_{zz}/De$, $\tau_{yz} = c_{yz}/De$ and $\tau_{yy} = (c_{yy} - 1)/De$; for the formulation in terms of the components of the conformation tensor, see Eqs. (A.1)-(A.4) in Appendix A.

In Eq. (2.9), the polymer viscosity ratio, $\eta$, appears:

$$\eta \equiv \frac{\eta_p^*}{\eta_s^* + \eta_p^*}. \qquad (2.14)$$

For $\eta=0$, i.e., in absence of the polymeric molecules, Eqs. (2.8)-(2.13) reduce to those for an incompressible Newtonian fluid. For $\eta=1$, the fluid is a pure polymeric material (UCM model), while for $0<\eta<1$ the fluid consists of a Newtonian solvent and a polymeric material (Oldroyd-B model). The dimensionless domain of definition of Eqs. (2.8)-(2.13) is:

$$\Omega = \{(z,y) \mid 0 < z < 1, -H(z) < y < H(z)\}.$$

The dimensionless form of the shape function is:

$$H(z) = \frac{1}{1 + (\Lambda - 1)z}, \quad 0 \leq z \leq 1. \qquad (2.15)$$

The case $\Lambda=1$ corresponds to a straight channel. In the entrance region ($z \leq 0$) $H(z)=1$, while in the exit region ($z \geq 1$), $H(z) = 1/\Lambda$. Thus, $H = H(z)$ is a piecewise smooth function which is continuous in the whole domain but not differentiable at $z=0$ and $z=1$. For $\Lambda > 1$, Eq. (2.15) describes a converging channel, and for $0 < \Lambda < 1$ an expanding channel; in the former case Λ will be referred to as the contraction ratio, as also mentioned before. An interesting property of the hyperbolic geometry is that the first derivative of $H$ with respect to $z$, $H'(z)$, is given in terms of $H$ only, i.e. $H'(z) = -(\Lambda-1)H^2(z)$. The latter leads to the recursive relation $H^{(k)}(z) = (-1)^k k!(\Lambda-1)^k H^{k+1}(z)$, $k = 0,1,2,3,\ldots$

The auxiliary dimensionless conditions (boundary conditions, total mass balance, symmetry conditions at the midplane, and the datum pressure) at the lubrication limit are:

$$V = U = 0 \quad \text{at} \quad y = H(z), 0 \leq z \leq 1, \tag{2.16a}$$

$$V = \frac{\partial U}{\partial y} = \tau_{yz} = 0 \quad \text{at} \quad y = 0, 0 \leq z \leq 1, \tag{2.16b}$$

$$\int_{-H(z)}^{H(z)} U(y,z)dy = 1 \quad \text{at} \quad 0 \leq z \leq 1, \tag{2.16c}$$

$$P(H(1),1) = 0. \tag{2.16c}$$

We close this section by reiterating that the classic lubrication equations for a viscoelastic fluid are, by construction, valid for large values of the Weissenberg number, clearly indicating a strong nonlinear rheological behavior of the flow. In the generalized Pipkin diagram (Pipkin 1986; Macosco 1994; Costanzo *et al.* 2024), it is important to recognize that the lubrication equations describe viscoelastic flow under conditions and rheological behavior of the fluid which corresponds to the upper part of the diagram (see Figure 1 in Costanzo *et al.* 2024). Regarding the magnitude of the Deborah number, acceptable values of *De* must be restricted to the left part of the Pipkin diagram, as in the upper-right region the rheological and flow conditions are such that the material behaves like an elastic solid, for which the Oldroyd-B model and other similar constitutive models are not valid. While it is challenging to determine the exact range of permissible Deborah numbers, it is evident that as the analysis approaches sufficiently larger *De* values, the rheological and flow conditions shift toward the upper-right region of the Pipkin diagram.

## 3. New set of lubrication equations

The solution procedure consists of three main steps and an assumption regarding the dependence of the unknown field variables on the spatial coordinates. Although the three steps of the solution procedure can be implemented altogether, it is instructive to report them separately and comment on the reasoning behind each step.

First, we introduce the streamfunction, $\Psi$, defined with the aid of the two velocity components:

$$U = \frac{\partial \Psi}{\partial y}, \quad V = -\frac{\partial \Psi}{\partial z}, \tag{3.1}$$

so that Eq. (2.8) is satisfied automatically. Therefore, when solving the final set of governing equations (asymptotically, exactly, or numerically) fluxes are satisfied with no error.

Second, we introduce new independent coordinates $(Y,Z)$ that map the varying boundaries of the flow domain into fixed ones:

$$Y = \frac{y}{H(z)}, \quad Z = z. \tag{3.2}$$

Thus, the domain of definition of the lubrication equations becomes $\bar{\Omega} = \{(Y,Z) \mid -1 < Y < 1, 0 < Z \leq 1\}$. Based on Eq. (3.2), the first order partial differential operators and the material derivative are transformed as follows:

$$\frac{\partial}{\partial y} = \frac{1}{H}\frac{\partial}{\partial Y}, \quad \frac{\partial}{\partial z} = \frac{\partial}{\partial Z} - \frac{YH'(Z)}{H(Z)}\frac{\partial}{\partial Y}, \quad \frac{D}{Dt} \equiv \frac{1}{H}\left(\frac{\partial \Psi}{\partial Y}\frac{\partial}{\partial Z} - \frac{\partial \Psi}{\partial Z}\frac{\partial}{\partial Y}\right). \tag{3.3}$$

Higher order derivatives can be readily obtained from the expressions provided in Eq. (3.3). The transformation presented in Eq. (3.2) is simple, straightforward, and convenient for both numerical and analytical purposes (Housiadas & Beris, 2023). Note that the use of the streamfunction and the new mapped coordinates was first introduced in the lubrication equations by Williams (1963). Williams studied steady Newtonian inertial flow in planar 2D channels and 3D axisymmetric pipes, both with variable walls.

Third, we introduce new components of the polymer extra-stress tensor, $\tau_{ss}, \tau_{ns}$, and $\tau_{nn}$, according to the linear expressions:

$$\tau_{ss} = H^4 \tau_{zz}, \quad \tau_{ns} = H^2 \tau_{yz} + (\Lambda - 1)Y H^4 \tau_{zz}, \quad \tau_{nn} = \tau_{yy} + 2(\Lambda - 1)Y H^2 \tau_{yz} + (\Lambda - 1)^2 Y^2 H^4 \tau_{zz} \tag{3.4}$$

This transformation was inspired by an expected property of the solutions presented recently by Housiadas & Beris (2023; 2024a). Indeed, both the eight-order asymptotic solution in terms of *De* and the numerical pseudospectral solution for $H^4 \tau_{zz}$, $H^2 \tau_{yz}$ and $\tau_{yy}$ revealed that these quantities are independent of the axial coordinate $Z$ (for $Z > 0.1$, approximately) in the range of the Deborah number investigated. A similar property was also revealed for the 3D axisymmetric case by Housiadas & Beris (2024d). The transformation given in Eq. (3.4) removes the discontinuity in $\tau_{yz}$ and $\tau_{yy}$ which is consequence of the jump of the slope of the shape function at the inlet (*Z*=0) of the hyperbolic section of the channel, as previously discussed by Housiadas & Beris (2023) (see subsection III.E in their paper). Eq. (3.4a) can be easily inverted to uniquely determine the original components of the polymer extra-stress tensor $\tau_{zz}, \tau_{yz}$, and $\tau_{yy}$, in terms of the new components $\tau_{ss}, \tau_{ns}$, and $\tau_{nn}$:

$$\tau_{zz} = \frac{\tau_{ss}}{H^4}, \quad \tau_{yz} = \frac{\tau_{ns} - (\Lambda - 1)Y\tau_{ss}}{H^2}, \quad \tau_{yy} = \tau_{nn} - 2(\Lambda - 1)Y\tau_{ns} + (\Lambda - 1)^2 Y^2 \tau_{ss} \quad (3.5)$$

Using Eqs. (3.1)-(3.4) into Eqs. (2.8)-(2.13) and after straightforward algebraic manipulations gives the final system of partial differential equations for $\Psi, \tau_{ss}, \tau_{ns}$ and $\tau_{nn}$:

$$-\frac{dP}{dZ}H^3 + (1-\eta)\frac{\partial^3 \Psi}{\partial Y^3} + \eta\left(3(\Lambda - 1)\tau_{ss} + \frac{\partial \tau_{ns}}{\partial Y} + \frac{1}{H}\frac{\partial \tau_{ss}}{\partial Z}\right) = 0 \quad (3.6)$$

$$\tau_{ss} + De\left(\frac{D\tau_{ss}}{Dt} + 2(\Lambda - 1)\tau_{ss}\frac{\partial \Psi}{\partial Y} - 2\tau_{ns}\frac{\partial^2 \Psi}{\partial Y^2} - 2\frac{\tau_{ss}}{H}\frac{\partial^2 \Psi}{\partial Z \partial Y}\right) = 0 \quad (3.7)$$

$$\tau_{ns} + De\left(\frac{D\tau_{ns}}{Dt} + 2(\Lambda - 1)\tau_{ns}\frac{\partial \Psi}{\partial Y} - \tau_{nn}\frac{\partial^2 \Psi}{\partial Y^2} + \frac{\tau_{ss}}{H}\left((\Lambda - 1)\frac{\partial \Psi}{\partial Z} + \frac{1}{H}\frac{\partial^2 \Psi}{\partial Z^2}\right)\right) = \frac{\partial^2 \Psi}{\partial Y^2} \quad (3.8)$$

$$\tau_{nn} + De\left(\frac{D\tau_{nn}}{Dt} + 2(\Lambda - 1)\tau_{nn}\frac{\partial \Psi}{\partial Y} + 2\frac{\tau_{nn}}{H}\frac{\partial^2 \Psi}{\partial Y \partial Z} + 2\frac{\tau_{ns}}{H}\left((\Lambda - 1)\frac{\partial \Psi}{\partial Z} + \frac{1}{H}\frac{\partial^2 \Psi}{\partial Z^2}\right)\right) = \\ -2\left((\Lambda - 1)\frac{\partial \Psi}{\partial Y} + \frac{1}{H}\frac{\partial^2 \Psi}{\partial Y \partial Z}\right) \quad (3.9)$$

where from this point onward we will use the modified pressure gradient term $G(Z) \equiv -P'(Z)H^3(Z)$ and the rationale for underlying certain terms in the equations will be explained further below. The accompanying boundary conditions for the streamfunction emerge from the no-slip and no-penetration conditions for velocity on the channel walls in conjunction with the symmetry of the flow field about the midplane:

$$\Psi(Y = \pm 1, Z) = \pm\frac{1}{2}, \quad \Psi(Y = 0, Z) = \frac{\partial^2 \Psi}{\partial Y^2}(Y = 0, Z) = \frac{\partial \Psi}{\partial Y}(Y = \pm 1, Z) = 0 \quad (3.10)$$

at any axial distance from the inlet, i.e. for $0 \leq Z \leq 1$. Obviously, all the Z-derivatives of the stream function at $Y = 0$ and $Y = \pm 1$ are zero, and the same holds for the Z-derivatives of $\partial \Psi / \partial Y\big|_{Y=\pm 1}$ and $\partial^2 \Psi / \partial Y^2\big|_{Y=0}$. Boundary conditions for the polymer extra-stress tensor at the inlet of the hyperbolic section of the channel are trivially derived from the analytical solution in the entrance region (see below in subsection 3.2).

The new set of lubrication equations can also be expressed in terms of the conformation tensor, by substituting $c_{ss} = De\,\tau_{ss}$, $c_{ns} = De\,\tau_{ns}$ and $c_{nn} = 1 + De\,\tau_{nn}$ into Eqs. (3.5)-(3.9). Since the relationship between the components of the extra-stress tensor and the conformation tensor is linear, the two formulations are equivalent. However, special care is required at the zero Deborah number limit as $\lim_{De \to 0} c_{zz} = \lim_{De \to 0} c_{yz} = 0$ and $\lim_{De \to 0} c_{yy} = 1$, but $\lim_{De \to 0} (c_{zz}/De) = \tau_{zz,N}$, $\lim_{De \to 0} (c_{yz}/De) = \tau_{yz,N}$ and $\lim_{De \to 0} (c_{yy} - 1)/De = \tau_{yy,N}$, where hereafter the subscript "*N*" denotes the corresponding Newtonian fluid solution. An important property of the conformation tensor is that it is positive definite (positive semidefinite at the lubrication limit); a violation of this property renders the results physically meaningless. Finally, we emphasize that the equations expressed in terms of the conformation tensor can be readily shown to be identical to those derived by Hinch, Boyko, and Stone (2024); see Appendix A for more details.

Even though the new set of governing equations is slightly more complicated than the original set of lubrication equations, they are much easier to handle either analytically, asymptotically, or numerically. Notice that for a straight channel, such as the entrance region (i.e., for $H = \Lambda = 1$), the mapped coordinates and the differential operators are identical to the original ones, i.e. $y \equiv Y, z \equiv Z$ and $\partial/\partial y \equiv \partial/\partial Y$, $\partial/\partial z \equiv \partial/\partial Z$, and the same holds for the transformed components of the polymer extra-tensor, i.e., $\tau_{ss} = \tau_{zz}$, $\tau_{ns} = \tau_{yz}$ and $\tau_{nn} = \tau_{yy}$. Therefore, Eqs. (3.5) – (3.9) can be used directly in the entrance region as well.

We proceed by studying some limiting cases of Eqs. (3.5)-(3.9) such as the Newtonian fluid for $\Lambda \geq 1$, the Oldroyd-B fluid for $\Lambda = 1$, and the exact analytical solution of the Oldroyd-B model for $\Lambda \geq 1$ at the channel walls (i.e., at $Y = \pm 1$), and on the midplane (i.e., at $Y = 0$). Note that both the Newtonian case for $\Lambda \geq 1$ and the Oldroyd-B fluid for $\Lambda = 1$ correspond to $De_m = 0$.

### *3.1. Newtonian solution*

For a Newtonian fluid ($De=De_m=0$) and $\Lambda \geq 1$, Eqs. (3.7)-(3.9) reduce to

$$\tau_{ss,N} = 0, \quad \tau_{ns,N} = \frac{\partial^2 \Psi_N}{\partial Y^2}, \quad \tau_{nn,N} = -2\left((\Lambda-1)\frac{\partial \Psi_N}{\partial Y} + \frac{1}{H}\frac{\partial^2 \Psi_N}{\partial Y \partial Z}\right). \tag{3.11}$$

Substituting Eqs. (3.11) into (3.6) gives:

$$0 = G_N + \frac{\partial^3 \Psi_N}{\partial Y^3}. \tag{3.12}$$

Solving Eq. (3.12) along with the appropriate boundary conditions (3.10), gives:

$$\Psi_N = \frac{3Y}{4} - \frac{Y^3}{4}, \quad G_N = \frac{3}{2}. \tag{3.13}$$

Finally, from Eqs. (3.11) and (3.13), the (Newtonian) extra stresses are found:

$$\tau_{ss,N} = 0, \quad \tau_{ns,N} = -\frac{3}{2}Y, \quad \tau_{nn,N} = -\frac{3}{2}(\Lambda-1)(1-Y^2) \quad (3.14)$$

We emphasize that Eqs. (3.13) and (3.14) are valid both in the entrance region (i.e., for Z<0, where $\Lambda = 1$) and in the varying section of the channel (i.e., for $0 \leq Z \leq 1$, where $\Lambda > 1$). At the inlet (Z=0) we observe that $\Psi_N$, $G_N$, $\tau_{ss,N}$ and $\tau_{ss,N}$ are continuous functions, whereas $\tau_{nn,N}$ exhibits a jump discontinuity, except at the walls of the channel (since $\tau_{nn,N}(Y=\pm 1)=0$). The discontinuity occurs due to the change in the slope of the shape function, and is interesting that affects only $\tau_{nn,N}$; the latter however, does not contribute in the modified pressure gradient, or the average pressure drop required to drive the flow in the channel at a constant flow rate.

The analytical solution for the Newtonian fluid, Eq. (3.13), can help us to extract some additional information for the original velocity components:

$$U_N = \frac{1}{H}\frac{\partial \Psi_N}{\partial Y} = \frac{3}{4H(Z)}(1-Y^2)$$
$$V_N = -\frac{\partial \Psi_N}{\partial Z} - (\Lambda-1)HY\frac{\partial \Psi_N}{\partial Y} = -(\Lambda-1)H(Z)\frac{3}{4}Y(1-Y^2) \quad (3.15)$$

Using the notation $f(Z \to 0^-) \equiv f^{(-)}$ and $f(Z \to 0^+) \equiv f^{(+)}$ for any dependent field variable $f$, Eq. (3.15) shows that $U_N^{(-)} = U_N^{(+)}$ and therefore $U$ is a continuous function throughout the entire length of the channel, whereas $V_N^{(-)} \neq V_N^{(+)}$, since $V_N^{(-)} = 0$ and $V_N^{(+)} = -3(\Lambda-1)HY(1-Y^2)/4$. This reveals that the original vertical velocity component exhibits a jump discontinuity at the inlet. In contrast, if the velocity components in the curvilinear coordinates $u_N = (1/H)(\partial \Psi_N / \partial Y)$ and $\upsilon_N = -\partial \Psi_N / \partial Z$ as employed by Hinch, Boyko & Stone (2024) (see Appendix A for more details), are used, the jump discontinuity is removed, and both $u_N$ and $\upsilon_N$ are continuous functions at the inlet of the hyperbolic section of the channel.

### 3.2. Oldroyd-B solution at the entrance region

The well-known exact analytical solution of the Oldroyd-B model in the entrance region (Z<0) can be found easily setting $H = \Lambda = 1$ and ignoring the Z-derivatives in Eqs. (3.6)-(3.9):

$$\Psi_S = \frac{3Y}{4} - \frac{Y^3}{4}, \quad G_S = \frac{3}{2}, \quad (3.16)$$

$$\tau_{ss,S} = 2De\left(\frac{\partial^2 \Psi_S}{\partial Y^2}\right)^2 = \frac{9}{2}De\,Y^2, \quad \tau_{ns,S} = \frac{\partial^2 \Psi_S}{\partial Y^2} = -\frac{3}{2}Y, \quad \tau_{nn,S} = 0 \qquad (3.17)$$

where the subscript "S" denotes the solution for the Oldroyd-B model in a straight channel. Eqs. (3.16) and (3.17) have been extensively used in the literature of non-Newtonian fluid mechanics for both analytical and computational purposes. Notice also that Eq. (3.16) is identical to Eq. (3.13); in other words the velocity profile and the pressure gradient in the entrance region of the channel are not affected by viscoelasticity ($\Psi_S = \Psi_N$ and $G_S = G_N$). The same holds for the *ns*- and *nn*-components of the polymer extra-stress tensor ($\tau_{ns,S} = \tau_{ns,N}$ and $\tau_{nn,S} = \tau_{nn,N}$) but not for the *ss*-components ($\tau_{ss,S} \neq \tau_{ss,N}$). Finally, it is worth noting that at the midplane the polymer molecules are unstretched (see Eq. (3.17) with *Y*=0) which is a consequence of the lubrication approximation, the symmetry of the flow about the midplane, and the fact that in the entrance region the channel is straight and the flow fully developed.

### 3.3. Exact solution of the Oldroyd-B model at the midplane

The exact analytical solution of the constitutive model at the midplane ($Y = 0$) has been derived for the 2D planar and 3D axisymmetric geometries by Housiadas & Beris (2023) and (2024d), respectively. Taking into account the conditions given in Eq. (3.10), Eqs. (3-7)-(3.9) reduce to:

$$T_{ss} + De\left(u\frac{dT_{ss}}{dZ} - 2T_{ss}\frac{du}{dZ} + 4(\Lambda - 1)\,u\,H\,T_{ss}\right) = 0, \qquad (3.18)$$

$$T_{ns} + De\,u\left(\frac{dT_{ns}}{dZ} + 2(\Lambda - 1)\,H\,T_{ns}\right) = 0, \qquad (3.19)$$

$$T_{nn} + 2\frac{du}{dZ} + De\left(u\frac{dT_{nn}}{dZ} + 2T_{nn}\frac{du}{dZ}\right) = 0, \qquad (3.20)$$

where $u(Z) \equiv \partial\Psi/\partial Y|_{Y=0}/H(Z)$, $T_{ss}(Z) \equiv \tau_{ss}(Y=0,Z)$, $T_{ns}(Z) \equiv \tau_{ns}(Y=0,Z)$ and $T_{nn}(Z) \equiv \tau_{nn}(Y=0,Z)$ have been used for brevity. Noting that $u(Z)$ is positive, the analytical solution of the first-order ODEs (3.18)-(3.20) is given with the aid of the exponentially decaying function, which is defined for any $De > 0$ as follows:

$$\varphi(Z) = \exp\left(-\frac{1}{De}\int_0^Z \frac{ds}{u(s)}\right), \qquad (3.21)$$

The solution is:

$$\tau_{ss}(0,Z) = \frac{\tau_{ss}(0,0)}{u^2(0)} u^2(Z)\varphi(Z) \qquad (3.22)$$

$$\frac{\tau_{ns}(0,Z)}{H^2(Z)} = \tau_{ns}(0,0)\varphi(Z) \qquad (3.23)$$

$$\tau_{nn}(0,Z) = \frac{1}{De}\left(-1 + \frac{\varphi(Z)}{u^2(Z)}\left(u^2(0)\left(De\,\tau_{nn}(0,0)+1\right) + \frac{1}{De}\int_0^Z \frac{u(s)}{\varphi(s)}ds\right)\right) \qquad (3.24)$$

For the derivation of Eqs. (3.22)-(3.24), $H(0)=1$ has been used, while the initial (in a Lagrangian sense) conditions for the polymer extra-stresses at $Z$=0 have been left unspecified. However, the exponential decay of the initial conditions is worth mentioning.

Using the exact solution of the Oldroyd-B model at the entrance region, i.e., Eqs. (3.16)-(3.17), we find $u(0) = 3/4$ and $\tau_{ss}(0,0) = \tau_{ns}(0,0) = \tau_{nn}(0,0) = 0$, and thus Eqs. (3.22)-(3.24) reduce to:

$$\tau_{ss}(0,Z) = \tau_{ns}(0,Z) = 0, \quad \tau_{nn}(0,Z) = \frac{1}{De}\left(-1 + \frac{\varphi(Z)}{u^2(Z)}\left(\frac{9}{16} + \frac{1}{De}\int_0^Z \frac{u(s)}{\varphi(s)}ds\right)\right) \qquad (3.25\text{a-c})$$

In terms of the components of the conformation tensor, Eq. (3.25a-c) gives:

$$c_{ss}(0,Z) = c_{ns}(0,Z) = 0, \quad c_{nn}(0,Z) = \frac{\varphi(Z)}{u^2(Z)}\left(\frac{9}{16} + \frac{1}{De}\int_0^Z \frac{u(s)}{\varphi(s)}ds\right) \qquad (3.26\text{a-c})$$

From Eq. (3.26c) we see that $c_{nn}(0,0)=1$, and is trivial to verify that $c_{nn}(0,Z)$ is strictly positive for any $Z \in [0,1]$ as well as it decreases monotonically with $Z$. Physically, $c_{ss}(0,Z)=0$ and $0 < c_{nn}(0,Z) < 1$ indicate that the polymer molecules on the midplane of the channel convected downstream by the fluid with velocity $u = u(Z)$ remain unstretched along the main flow direction but are compressed in the direction towards the wall and perpendicular to the midplane.

### 3.4. *Exact solution of the Oldroyd-B model at the walls*

A key component of the theoretical analysis of the flow, also essential for deriving high-accuracy numerical results, is the exact analytical solution of the constitutive model along the channel walls. This solution has been previously derived for an arbitrary shape function and discussed extensively by Housiadas & Beris (2023; 2024a–c) for the 2D planar case, and by Housiadas & Beris (2024d) and Sialmas & Housiadas (2025) for the 3D axisymmetric case. Indeed, by denoting the shear stress at the wall as $\Gamma_\pm(Z) \equiv \partial^2\Psi/\partial Y^2\big|_{Y=\pm 1}$ (where $\Gamma_+(Z) = -\Gamma_-(Z)$ due to the antisymmetry with respect to the midplane), and using the conditions given in Eq. (3.10), Eqs. (3.7)-(3.9) can also be solved analytically to yield:

$$\tau_{ss}(\pm 1, Z) = 2De\,\Gamma_\pm^2(Z), \quad \tau_{ns}(\pm 1, Z) = \Gamma_\pm(Z), \quad \tau_{nn}(\pm 1, Z) = 0. \tag{3.27}$$

Equation (3.27) provides the exact solution for the modified polymer extra-stress components in the special case of a hyperbolic channel. It fully aligns with the general solution derived by Housiadas & Beris (2023) for an arbitrary shape function, expressed in terms of the original components of the polymer extra-stress tensor [see Eq. (62a, b, c) in Housiadas & Beris (2023)]. The second expression in Eq. (3.27) highlights a crucial aspect of the flow, which has been entirely overlooked by other researchers: the Oldroyd-B model predicts that the modified shear component of the polymer extra-stress tensor is exactly equal to the pure viscous shear force. For this reason, Eq. (3.27) is central to accurately determining the solution near the walls and correctly evaluating the average pressure drop required to maintain a constant flow rate, using a total force balance on the channel (see Section 4 below).

Comparing the solution at the walls in the entrance region (Eq. (3.17)) with the solution at the walls in the varying region (Eq. (3.27)), we observe that $\tau_{ss}^{(-)}(Y = \pm 1) = \tau_{ss}^{(+)}(Y = \pm 1)$, $\tau_{ns}^{(+)}(\pm 1) = \tau_{ns}^{(+)}(\pm 1)$, and $\tau_{nn}^{(+)}(\pm 1) = \tau_{nn}^{(+)}(\pm 1)$. These results arise from the continuity of the streamfunction in the Z direction; this implies that $\Psi^{(-)}(Y) = \Psi^{(+)}(Y)$, and the same holds for its Y-derivatives. Therefore, the discontinuity of the $\tau_{yz}$ and $\tau_{yy}$ at the inlet of the hyperbolic section of the channel is removed by using the new components of the polymer extra-stress tensor $\tau_{ss}$, $\tau_{ns}$ and $\tau_{nn}$.

In Figure 2, we present a sketch of the upper part of the channel, including all regions, namely the entrance region ($Z < 0$), the hyperbolic section ($0 \leq Z \leq 1$), and the exit region $Z > 1$. The conditions along the wall (Y=1) and on the midplane (Y=0) are given in terms of the mapped coordinates, the streamfunction, and the new components of the polymer extra-stress tensor. The analytical solution at the entrance region is also shown. Unlike the original formulation, all the field variables are continuous both along the wall and the midplane. However, the modified pressure gradient $G$ at the inlet ($Z = 0$) exhibits a jump discontinuity due to the change in the slope of the shape function in conjunction with the primary normal viscoelastic stresses. Unlike the discontinuity in the original stresses, this pressure gradient discontinuity cannot be avoided or removed. Its origin, magnitude, and effect on the flow are described in the next subsection.

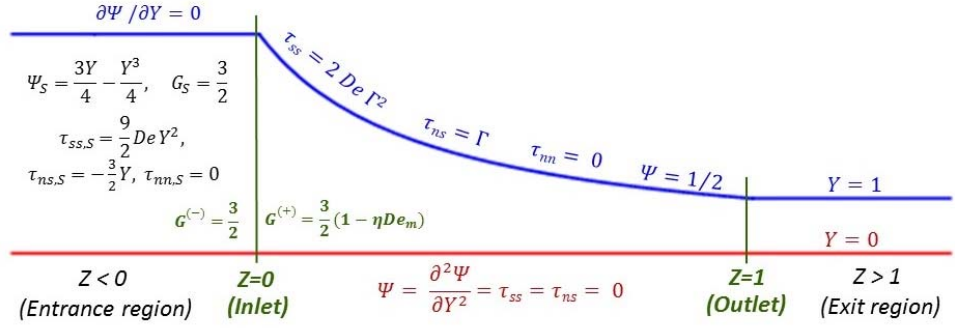

**Figure 2:** Conditions along the wall (*Y=1*) and on the midplane (*Y=0*) in terms of the mapped coordinates, the streamfunction, and the new transformed components of the polymer extra-stress tensor. All conditions hold for all regions of the channel. Along the wall, $\Gamma = \frac{\partial^2 \Psi}{\partial Y^2}(Y=1, Z)$. The analytical solution at the entrance region (Z<0) is also given (indicated by the "S" subscript). At *Y=0* and *Y=1*, all components of the polymer extra-stress tensor are continuous functions of *Z*, whereas the pressure gradient *G* exhibits a jump discontinuity at *Z=0*.

### *3.5. Pressure-gradient discontinuity at the inlet*

We start by noting that in the entrance region $Z \leq 0$ all field variables are known and continuous functions (see Eqs. (3.16)-(3.17)). Their *Y*-derivatives are also known for $Z \leq 0$, as well as their *Z*-derivatives for $Z < 0$. The shape function $H$ is similarly known and continuous across all regions of the channel (entrance, varying, and exit regions), with $H(0^-) = H(0^+) = 1$ and $H'(0^-) = 0$ and $H'(0^+) = -(\Lambda - 1)$, exhibiting a jump discontinuity in its slope at the inlet of the hyperbolic section.

We proceed, by evaluating Eq. (3.6) at $Z \to 0^-$ (where $\Lambda = 1$):

$$G^{(-)} + (1-\eta)\frac{\partial^3 \Psi^{(-)}}{\partial Y^3} + \eta\left(\frac{\partial \tau_{ns}^{(-)}}{\partial Y} + \left(\frac{\partial \tau_{ss}}{\partial Z}\right)^{(-)}\right) = 0 \qquad (3.28)$$

Using that $\partial^3 \Psi^{(-)}/\partial Y^3 = -3/2$, $\partial \tau_{ns}^{(-)}/\partial Y = -3/2$, and $(\partial \tau_{ss}/\partial Z)^{(-)} = 0$ in Eq. (3.28) gives $G^{(-)} = G_S = 3/2 > 0$. The latter represents a favorable pressure gradient, meaning that in the entrance region, the pressure decreases downstream, acting to maintain the flow at a constant speed.

Similarly, evaluating Eq. (3.6) at $Z \to 0^+$, where $\Lambda > 1$, yields:

$$G^{(+)} + (1-\eta)\frac{\partial^3 \Psi^{(+)}}{\partial Y^3} + \eta\left(3(\Lambda-1)\tau_{ss}^{(+)} + \frac{\partial \tau_{ns}^{(+)}}{\partial Y} + \left(\frac{\partial \tau_{ss}}{\partial Z}\right)^{(+)}\right) = 0 \qquad (3.29)$$

Using $\partial^3 \Psi^{(+)} / \partial Y^3 = -3/2$, $\tau_{ss}^{(+)} = 9DeY^2/2$, $\partial \tau_{ns}^{(+)} / \partial Y = -3/2$ and $(\partial^2 \Psi / (\partial Y \partial Z))^{(+)} = 0$, the last term in Eq. (3.29) is found from Eq. (3.7) at the limit $Z \to 0^+$ as $(\partial \tau_{ss} / \partial Z)^{(+)} = -9De_m Y^2$. Substituting all quantities in Eq. (3.29) and integrating from $Y$=0 to $Y$=1 yields:

$$G^{(+)} = \frac{3}{2}(1 - \eta De_m) \tag{3.30}$$

Since $G^{(-)} \ne G^{(+)}$, $G$ is discontinuous at $Z$=0, with a jump $[\![G]\!] = G^{(+)} - G^{(-)} = -3\eta De_m / 2$ in the direction of increasing $Z$. Therefore, the sudden change of the slope of the geometry at the inlet and the normal extra-stresses due to the polymer viscoelasticity induces the jump discontinuity in the pressure gradient.

Eq. (3.30) reveals a very interesting and unexpected feature for a pressure-driven slow flow. For $1 - \eta De_m > 0$ the pressure gradient at the inlet remains favorable to maintain the constant flow rate as the fluid enters the contraction. However, when $1 - \eta De_m < 0$, the pressure gradient changes sign at the inlet and becomes negative. Therefore, the fluid suddenly experiences an adverse pressure gradient; for instance, for $\eta = 4/10$ and $De_m = 4$, $G^{(-)} = 1.5$ and $G^{(+)} = -0.9$. It is unclear whether, for a slow and steady-state flow in a confined geometry, such a sudden adverse pressure gradient can be considered physical or realistic, even if it is localized to a specific cross section in the flow domain. There is no doubt however that such a phenomenon must be restricted in a boundary layer near the inlet, $0 < Z \ll 1$, and the pressure gradient must eventually recover a positive value since an increase of the pressure downstream is not physically acceptable (indeed this is confirmed by the full numerical simulations). Although, the region $1 - \eta De_m < 0$ will be investigated below too, it is reasonable to assume that the threshold $\eta De_m = 1$ indicates transition to viscoelastic instabilities or the loss of validity of the classic lubrication approximation for an Oldroyd-B fluid.

## 4. General formulas for pressure gradient and average pressure-drop

A major quantity of engineering interest is the average pressure drop, $\Delta \Pi$, required to maintain a constant flowrate through the channel. At the classic lubrication limit, and using the definition of the modified pressure gradient term $G(Z)$, $\Delta \Pi$ is calculated generally as:

$$\Delta \Pi = -\int_0^1 P'(Z) dZ = \int_0^1 \frac{G(Z)}{H^3(Z)} dZ. \tag{4.1}$$

For a Newtonian fluid (i.e., for $De = De_m = 0$), $G(Z) = G_N(Z) = 3/2$ and Eq. (4.1) gives:

$$\Delta \Pi_N = \frac{3}{2} \int_0^1 \frac{dZ}{H^3(Z)} = \frac{3}{8}(1+\Lambda)(1+\Lambda^2).  \tag{4.2}$$

From Eqs. (4.1)-(4.2) one can derive the reduced average pressure drop $\Delta\Pi / \Delta\Pi_N$ which is presented further below to investigate the effect of viscoelasticity. The modified pressure gradient $G$ is found from the momentum balance, Eq. (3.6), after integration with respect to $Y$ from the midplane ($Y=0$) to the wall ($Y=1$), using the symmetry and boundary conditions at the midplane and wall, respectively, and taking into account the exact analytical solution of the Oldroyd-B model at the wall:

$$G + \frac{\partial^2 \Psi}{\partial Y^2}\bigg|_{Y=1} + \eta \left( 3(\Lambda - 1) I + \frac{1}{H} \frac{dI}{dZ} \right) = 0, \quad I(Z) = \int_0^1 \tau_{ss}(Y,Z) dY \tag{4.3}$$

At this point, it is essential to emphasize the significance of the general exact solution of the Oldroyd-B model at the walls, as presented in Eq. (3.27), expressed in terms of the modified shear rate at the wall, $\partial^2 \Psi / \partial Y^2 \big|_{Y=1}$. This solution implies that integrating the terms corresponding to the shear stresses, from the midplane to the wall, yields precisely $\partial^2 \Psi / \partial Y^2 \big|_{Y=1}$, i.e.

$$\int_0^1 \left( (1-\eta) \frac{\partial^3 \Psi}{\partial Y^3} + \eta \frac{\partial \tau_{ns}}{\partial Y} \right) dY = \frac{\partial^2 \Psi}{\partial Y^2}\bigg|_{Y=1} \tag{4.4}$$

Substituting $G$ from Eq. (4.3) into Eq. (4.1) and simplifying the result, yields the final expression for the average pressure drop:

$$\Delta \Pi = \underbrace{\int_0^1 -\frac{1}{H^3} \frac{\partial^2 \Psi}{\partial Y^2}\bigg|_{Y=1} dZ}_{\dot{\gamma}_w} + \underbrace{\eta \left( I(0) - \Lambda^4 I(1) + (\Lambda - 1) \int_0^1 \frac{I(Z)}{H^3(Z)} dZ \right)}_{\tau} \tag{4.5}$$

Eq. (4.5) reveals that when the exact solution of the Oldroyd-B model at the wall is used, the average pressure drop is given at the simplest possible form, i.e., $\Delta\Pi = \dot{\gamma}_w + \tau$, where $\dot{\gamma}_w$ is the contribution due to the viscous shear forces at the wall, and $\tau$ is the viscoelastic bulk contribution arising exclusively from the principal normal component, $\tau_{ss}$, of the polymer extra-stress tensor. Eq. (4.5) is the total force balance on the flow system at the classic lubrication limit. It has been previously derived using the original field variables that appear in Eqs (2.8)-(2.13) by Housiadas & Beris (2023; 20224a-c) for the 2D planar geometric configuration, as well as by Housiadas & Beris (2024d) and Sialmas & Housiadas (2025) for the 3D axisymmetric case in a cylindrical pipe. Alternatively, the procedure by Hinch, Boyko & Stone (2024) can be used to determine the average pressure drop.

Another formula for $\Delta \Pi$ based on the total mechanical energy of the system can also be derived as previously done by Housiadas & Beris (2024c) for the 2D planar case, and by Housiadas & Beris (2024d) and Sialmas & Housiadas (2025) for the 3D axisymmetric case. Briefly, Eq. (2.9) is multiplied with the main velocity component $U$ and is integrated over a cross-section of the channel. The resulting equation is integrated over the entire length of the hyperbolic section, while the terms that arise due to the viscoelasticity of the fluid are further simplified by means of Eq. (2.11), i.e., the component of the Oldroyd-B model which corresponds to the zz-component of the original polymer extra-stress tensor. Thus, the following expression is derived:

$$\Delta \Pi = \Phi_V + \Phi_{el} + W \quad (4.6)$$

where $\Phi_V$ is the pure viscous dissipation, $\Phi_{el}$ is the elastic bulk contribution, and $W$ is the work done by the elastic forces. Using the symbol $\Delta(f) := f(z=0) - f(z=1)$, these quantities are given in terms of the original coordinates and field variables as:

$$\Phi_V = 2(1-\eta) \int_0^1 \int_0^H \left( \frac{\partial U}{\partial y} \right)^2 dy dz, \quad \Phi_{el} = \frac{\eta}{De} \int_0^1 \int_0^H \tau_{zz} dy dz, \quad W = \eta \Delta \left( \int_0^H \tau_{zz} U dy \right) \quad (4.7)$$

or, in terms of the mapped coordinates and new field variables introduced in Section 3, as:

$$\Phi_V = 2(1-\eta) \int_0^1 \frac{1}{H^3} \left( \int_0^1 \left( \frac{\partial^2 \Psi}{\partial Y^2} \right)^2 dY \right) dZ, \quad \Phi_{el} = \frac{\eta}{De} \int_0^1 \frac{1}{H^3} \left( \int_0^1 \tau_{ss} dY \right) dZ, \quad W = \eta \Delta \left( \frac{1}{H^4} \int_0^1 \frac{\partial \Psi}{\partial Y} \tau_{ss} dY \right) \quad (4.8)$$

Eq. (4.8) has also been confirmed independently utilizing the same procedure on Eqs. (3.6) and (3.7). Eqs. (4.7) and (4.8) are equivalent to the corresponding equation for the mechanical energy of the flow under steady state by Hinch, Boyko & Stone (2024). It is also clear that $\Phi_V$ is strictly positive, and therefore purely dissipative, while for any $De > 0$, $\Phi_{el}$ is also dissipative because for the Oldroyd-B model $\tau_{ss}$ remains positive throughout the flow domain. The sign of $W$, however, is unclear.

## 5. Solutions for the Oldroyd-B model in the varying region ( $\Lambda > 1$ )

To solve Eqs. (3.6)–(3.10) governing the flow in the hyperbolic section of the channel, we develop three methods. The first is a high-order asymptotic technique, similar to that previously developed by Housiadas & Beris (2023; 2024a,c,d). The second method is a high-accuracy pseudospectral numerical method. The third method is a new exact analytical solution of the Oldroyd-B model, combined with the numerical solution of the momentum balance as this is simplified according to the exact solution.

The new exact solution represents the main and most significant contribution of this work, while the first two methods are employed to evaluate its accuracy and range of applicability. To ensure clarity, we emphasize that the term "exact" signifies that when the solution is substituted into Eqs. (3.7)–(3.9), turns them into identities.

### 5.1. High-order asymptotic solution for weakly viscoelastic fluids

For small values of the Deborah number, a regular perturbation scheme is assumed:

$$F(Y,Z) \approx F_N(Y) + \eta \sum_{k=1}^{\infty} F_k(Y,Z) De^k, \quad 0 < De \ll 1, \quad (5.1)$$

where $F = \Psi, \tau_{ss}, \tau_{ns}$, and $\tau_{nn}$. The series are substituted into (3.6)-(3.10) yielding a sequence of equations which are solved analytically up to $O(De^8)$ with the aid of the *MATHEMATICA* software (Wolfram 2023). The $O(1)$ equations correspond to those for Newtonian fluid, while the effect of viscoelasticity is incorporated through the higher-order terms. With the aid of the solution of the Oldroyd-B model for a straight channel, Eqs (3.16)-(3.17), and recalling that $\tau_{ss,S}(Y) = O(De)$, the solution up to second order is:

$$\Psi \approx \Psi_S(Y) + \frac{27}{560} \eta De_m^2 (-2 + 3Y^2 - Y^6) Y, \quad (5.2a)$$

$$\tau_{ss} \approx \tau_{ss,S}(Y) \left(1 - \frac{9}{2}(1-Y^2) De_m\right), \quad (5.2b)$$

$$\tau_{ns} \approx \tau_{ns,S}(Y) \left(1 - 3(1-Y^2) De_m + \frac{27}{140}\left(3\eta + 35(1-2Y^2) + 7(5+\eta)Y^4\right) De_m^2\right), \quad (5.2c)$$

$$\tau_{nn} \approx \tau_{nn,S}(Y) \left(1 - \frac{3}{2}(1-Y^2) De_m - \frac{9}{140}\left(2\eta - 35 + 7(10-\eta)Y^2 - 7(5+\eta)Y^4\right) De_m^2\right) \quad (5.2d)$$

The high-order asymptotic solution revealed that at any order of the approximation, all the field variables are independent of the *Z*-coordinate. Additionally, $\Psi$ and $\tau_{ns}$ are odd functions of *Y*, and $\tau_{ss}$ and $\tau_{nn}$ are even functions of *Y*. We also confirmed the absence of an $O(De)$ term for the streamfunction in agreement with the theorem of Tanner & Pipkin (1969), as previously also found for the solution of the original set of lubrication equations by Boyko & Stone (2022) and Housiadas & Beris (2023). Finally, using Eq. (3.5) to restore the original components of the polymer extra-stress tensor $\tau_{zz}, \tau_{yz}$ and $\tau_{yy}$, we verified that the solution derived here up to $O(De^8)$ fully agrees with that previously obtained by Housiadas & Beris (2023; 2024a,c).

The independence of the high-order asymptotic solution from *Z* demonstrates that the regular perturbation expansion in terms of $De$ is incapable of resolving the underlying terms in the lubrication equations at any order of approximation. This observation leads us to the following question: Is there a more general solution to Eqs. (3.6)-(3.10) that does not depend on *Z*? This issue is investigated and addressed in subsection 5.3, giving the main and most important result of this paper.

*5.2. Numerical solution of the new set of lubrication equations*

The numerical solution of the original lubrication equations, Eqs. (2.8)–(2.13), has been presented by Housiadas & Beris (2023) employing a variety of methods such as finite difference and pseudospectral schemes. Here we solve the new set of lubrications equations, Eqs. (3.6)–(3.10), by developing an substantially improved version of the previous pseudospectral code. A brief description of the numerical scheme is provided below, focusing on its main features.

Each dependent field variable, $f = f(Y,Z)$ where $f = \Psi, \tau_{ss}, \tau_{ns}$ and $\tau_{nn}$, is represented as a finite sum of Chebyshev orthogonal polynomials, $f(Y,Z) \approx f_{[M]}(Y,Z) = \sum_{k=0}^{M} \hat{f}_k(Z) T_k(Y)$ where $\hat{f}_k(Z)$ are the corresponding spectral coefficients and $T_k(Y)$ the Chebyshev polynomials defined over the domain $-1 \leq Y \leq 1$. Due to the symmetries of the flow, only the odd spectral coefficients contribute for $\Psi$ and $\tau_{ns}$, while only the even spectral coefficients contribute for $\tau_{ss}$ and $\tau_{nn}$. Thus, all flow symmetries are satisfied exactly by construction. Notably, due to the formulation in terms of $\Psi$, the fluxes are satisfied with zero error. At the midplane and channel walls, the exact solution of the Oldroyd-B model is imposed on the modified polymer extra-stress components, i.e., Eq. (3.25a-c) and (3.27), respectively. The calculations in the transformed vertical coordinate are performed pseudospectrally at the Gauss-Lobatto collocation points, $Y_j = -\cos(\pi j / M)$, $j = 0,1,2,...,M$, where *M* varies from 4 in the Newtonian case to 28 in the most demanding cases (see Appendix B for more details). Additionally, the convective terms in the constitutive model, Eqs. (3.7)–(3.9), are computed in conservative form (see Appendix A) to enhance both accuracy—by minimizing aliasing errors—and stability in the calculations.

The integration of Eqs. (3.6)–(3.9) in the Z-direction is performed as an initial value problem using the fully implicit A(0)-stable, first-order accurate backward differentiation formula. At the inlet of the hyperbolic section of the channel (Z=0), the exact analytical solution of the Oldroyd-B model, given by Eqs. (3.16)-(3.17), is imposed. At each cross-section of the channel, i.e. for any $Z_j = j/N, j = 1, 2, ..., N$, where $N$ represents the number of grid points along the Z-direction, the solution is found in a fully implicit fashion. Thus, at each Z-step, the unknown values of the field variables at the collocation points and the pressure gradient term *G* are determined. To this end, a Newton scheme is implemented which typically converges quadratically (in full accordance to the theory of Newton's scheme) within three iterations, with an absolute error criterion of $10^{-13}$-$10^{-12}$. Last, the spectral coefficients $\hat{\Psi}_k(Z_j), \hat{\tau}_{ss,k}(Z_j), \hat{\tau}_{ns,k}(Z_j)$ and $\hat{\tau}_{nn,k}(Z_j)$, $k = 0,1,2,...,M$ are calculated in order to monitor their magnitude and confirm the degree of resolution of the field variables.

The spectral accuracy and performance of the pseudospectral code are verified through four test cases with increasing complexity (see Appendix B for more details). Additionally, we note that two versions of the new pseudospectral code have been developed: the first solves for the polymer extra-stress components, Eqs. (3.6)–(3.9), while the second solves for the components of the conformation tensor (see Eqs. (A.5)-(A.8) in Appendix A). No advantages have been identified for either version, except that the version solving for the polymer extra-stress tensor can also handle the zero Deborah number limit.

### 5.3. An exact solution

Assuming that $\tau_{ss}(Y,Z) \approx \tilde{\tau}_{ss}(Y)$, $\tau_{ns}(Y,Z) \approx \tilde{\tau}_{ns}(Y)$, $\tau_{nn}(Y,Z) \approx \tilde{\tau}_{nn}(Y)$, $\Psi(Y,Z) \approx \tilde{\Psi}(Y)$ and $G(Z) \approx \tilde{G} = \text{constant}$ eliminates all the underlying terms in Eqs. (3.6)-(3.9). This assumption implies that a similarity solution to the lubrication equations is sought, with *Y* serving as the similarity variable. Williams (1963) followed the same procedure to find similarity solutions for the lubrication equation governing the flow of a Newtonian fluid, including fluid inertia, in confined geometries. Note that the elimination of the *Z*-derivatives from Eqs. (3.6)-(3.9) is not the same as the elimination of the *z*-derivatives from Eqs. (2.8)-(2.13), as can easily be confirmed from Eq. (3.3). It also implies that the convective derivative of the modified extra-stress components in terms of the new coordinates $(Y,Z)$ is zero, i.e. $Df/Dt = 0$ where $f = \tilde{\tau}_{ss}$, $\tilde{\tau}_{ns}$ and $\tilde{\tau}_{nn}$ (see Eq. (3.3)). On the contrary, the convective derivative of the original components of the extra-stress tensor are:

$$\frac{D\tau_{zz}}{Dt} = 4(\Lambda-1)\tilde{\Psi}'(Y)\frac{\tilde{\tau}_{ss}(Y)}{H^4(Z)}, \quad \frac{D\tau_{yz}}{Dt} = 2(\Lambda-1)\tilde{\Psi}'(Y)\frac{\tilde{\tau}_{ns}(Y) - Y(\Lambda-1)\tilde{\tau}_{ss}(Y)}{H^2(Z)}, \quad \frac{D\tau_{yy}}{Dt} = 0 \quad (5.3)$$

Regarding the streamfunction, due to its construction-definition, its convective derivative is zero at any coordinate system, $D\Psi/Dt = 0$, as, for instance, can be verified in the original coordinates $(y,z)$ using Eq. (3.1) and the definition of the convective derivative, or in the mapped coordinates $(Y,Z)$ using Eq. (3.3).

Under the assumption made at the beginning of this subsection, Eqs. (3.7)-(3.9) can be solved analytically to provide a unique solution for the modified polymer extra-stress components, valid theoretically for arbitrary values of $De$ and $De_m$:

$$\tilde{\tau}_{ss}(Y) = \frac{2De\,\tilde{\Psi}''^2}{(1+2De_m\,\tilde{\Psi}')^3}, \quad \tilde{\tau}_{ns}(Y) = \frac{\tilde{\Psi}''}{(1+2De_m\,\tilde{\Psi}')^2}, \quad \tilde{\tau}_{nn}(Y) = -\frac{2(\Lambda-1)\tilde{\Psi}'}{1+2De_m\,\tilde{\Psi}'} \quad (5.4\text{a-c})$$

where the prime denotes differentiation with respect to the mapped vertical coordinate Y. From Eq. (5.4a-c) we can also easily derive the corresponding exact solution for the conformation tensor components:

$$\tilde{c}_{ss}(Y) = \frac{2De^2\,\tilde{\Psi}''^2}{(1+2De_m\,\tilde{\Psi}')^3}, \quad \tilde{c}_{ns}(Y) = \frac{De\,\tilde{\Psi}''}{(1+2De_m\,\tilde{\Psi}')^2}, \quad \tilde{c}_{nn}(Y) = \frac{1}{1+2De_m\,\tilde{\Psi}'} \quad (5.5\text{a-c})$$

We note that in a lubrication flow, which by definition is a shear-dominated flow, it is of crucial importance that Eq. (5.4a-c), or Eq. (5.5a-c), satisfy the exact solution of the Oldroyd-B model at the walls of the channel. Indeed, at the walls ($Y = \pm 1$), $\tilde{\Psi}'(\pm 1) = 0$ due to the no-slip condition, and thus Eq. (5.4a-c) reduce to $\tilde{\tau}_{ss}(\pm 1) = 2De\,(\tilde{\Psi}''(\pm 1))^2$, $\tilde{\tau}_{ns}(\pm 1) = \tilde{\Psi}''(\pm 1)$ and $\tilde{\tau}_{nn}(\pm 1) = 0$, respectively. Comparing with Eq. (3.27) reveals that the new solution fully agrees and is consistent with the exact solution of the Oldroyd-B at the walls. However, and as expected, the similarity solution (Eqs. (5.4a-c) or Eq. (5.5a-c)) does not satisfy any condition at the inlet of the hyperbolic section of the channel.

Likewise, ignoring the Z-derivative in the momentum balance reduces Eq. (3.6) to:

$$\tilde{G} + (1-\eta)\tilde{\Psi}'''(Y) + \eta\big(3(\Lambda-1)\tilde{\tau}_{ss}(Y) + \tilde{\tau}'_{ns}(Y)\big) = 0 \quad (5.6)$$

Substituting $\tilde{\tau}_{ss}(Y)$ and $\tilde{\tau}_{ns}(Y)$ from Eq. (5.4b,c) in Eq. (5.6) gives a third-order ODE for the streamfunction:

$$\tilde{G}+(1-\eta)\tilde{\Psi}'''+\eta\left(\frac{2De_m\tilde{\Psi}''^2}{(1+2De_m\tilde{\Psi}')^3}+\frac{\tilde{\Psi}'''}{(1+2De_m\tilde{\Psi}')^2}\right)=0 \tag{5.7}$$

The constant pressure gradient $\tilde{G}$ is determined by integrating Eq. (5.6) with respect to $Y$, from the midplane ($Y=0$) to the wall ($Y=1$) and taking into account the boundary and symmetry conditions for the streamfunction as well as the exact solution of the Oldroyd-B model at the wall. Alternatively, Eq. (4.3) can be used directly. In both cases, one finds:

$$\tilde{G}+\tilde{\Psi}''(1)+3\eta(\Lambda-1)\int_0^1\tilde{\tau}_{ss}(Y)dY=0 \tag{5.8}$$

However, Eqs. (5.7)-(5.8) are strongly nonlinear and cannot be solved analytically to determine $\tilde{\Psi}=\tilde{\Psi}(Y)$ and $\tilde{G}$. Thus, Eqs. (5.7)-(5.8) are solved semi-numerically using a fully spectral method as opposed to the pseudospectral code used for the simulation of the full set of lubrication equations; the fully spectral code resolves the streamfunction almost down to machine accuracy and calculates the constant modified pressure gradient.

First, we note that the case $De_m=0$ is trivial because it corresponds to a pure Newtonian fluid ($De=0, \Lambda\geq 1$) or to the flow of an Oldroyd-B fluid in a straight channel ($De>0, \Lambda=1$). In both cases, Eq. (3.12), $\tilde{G}+\tilde{\Psi}'''=0$, is recovered from which we find the polymer extra-stresses, i.e. Eqs. (3.14) and (3.17).

We proceed for $De_m>0$ by defining $F(Y):=1+2De_m\tilde{\Psi}'(Y)$, $\hat{G}:=2De_m\tilde{G}/(1-\eta)$ and $c:=\eta/(1-\eta)\equiv\eta_p^*/\eta_s^*$. Thus, Eq. (5.7) reduces to a 2$^{nd}$-order ODE for $F$:

$$(\hat{G}+F'')F^3+c(F'^2+FF'')=0, \quad 0<Y<1 \tag{5.9a}$$

accompanied with the auxiliary conditions:

$$F(1)=1, \quad F'(0)=0, \quad \int_0^1 F(Y)dY=1+De_m \tag{5.9b}$$

The first expression in Eq. (5.9b) is the no-slip condition at the wall, the second corresponds to the symmetry with respect to the midplane, and the last is the constant flow-rate at each cross section of the channel due to fluid's incompressibility (see Eq. (2.16c) after introducing the mapped coordinates $(Y,Z)$). Obviously, $F$ depends on $Y$ and the dimensionless parameters $De_m$ and $\eta$ (or $c$).

The unknown function $F$ is approximated by $M$ even Legendre orthogonal polynomials as $F(Y)\approx F_{[M]}(Y)=\sum_{k=0}^{M-1}\hat{F}_{2k}P_{2k}(Y)$ where $\hat{F}_{2k}$ are the corresponding spectral coefficients [Hesthaven *et al.* 2007]. Due to the orthogonality property of the Legendre polynomials, $\int_{-1}^{1} P_k(Y)dY = 0$ for $k \geq 1$ and application of the integral condition given in Eq. (5.9b) determines $\hat{F}_0 = 1 + De_m$. Furthermore, by applying the first condition given in Eq. (5.9b) gives $\hat{F}_2 = -De_m - \sum_{k=2}^{M-1} \hat{F}_{2k}$, while the second condition in Eq. (5.9b) is satisfied automatically since $P'_{2k}(0) = 0$ for all positive values *k*. Therefore, all conditions in Eq. (5.9b) are satisfied by construction, assuming that:

$$F_{[M]}(Y) = 1 + \frac{3}{2} De_m (1 - Y^2) + \sum_{k=2}^{M-1} \hat{F}_{2k} P_{2k}(Y), \quad M \geq 2 \tag{5.10}$$

where $P_2(Y) = (-1 + 3Y^2)/2$ has been used. The case $M = 2$ gives $F_{[2]}(Y) = 1 + 3De_m(1-Y^2)/2$ and is the lowest-order approximation which in fact corresponds to the Newtonian velocity profile ($F_{[2]}(Y) \equiv F_N(Y)$). Substituting $F_{[2]}(Y)$ in Eq. (5.9a), and integrating with respect to *Y* from the lower (*Y*=-1) to the upper wall (*Y*=1), gives the lowest approximation for the modified pressure gradient term $\hat{G}_{[2]}$:

$$\frac{(1-\eta)}{2De_m} \hat{G}_{[2]} = \tilde{G}_{[2]} = \frac{3}{2}\left(1 - \eta + \frac{\eta}{1 + 3De_m + 126De_m^2/35 + 54De_m^3/35}\right) \tag{5.11}$$

For $M > 2$ the unknown modified pressure gradient and coefficients of $F_{[M]}(Y)$, i.e. $\hat{G}_{[M]}, \hat{f}_4, \hat{f}_6, ..., \hat{f}_{2(M-1)}$, are computed by weighting Eq. (5.9a) with $P_{2k}(Y)$, $k = 0, 1, 2, ..., M-2$. The resulting equations are integrated analytically with respect to *Y* from -1 to 1, using the *MATHEMATICA* software, i.e. a fully spectral Galerkin-type method is implemented. This procedure results in $M - 1$ non-linear algebraic equations, the solution of which is found using the build-in routine "*FindRoot*" of the same software (the case *M*=2 is used as an initial guess). Increasing *M*, i.e., considering more terms in the Legendre series for $F$, and implementing the algorithm described above results in a very fast convergence for $F$ and $\hat{G}$. We choose $\eta = 0.4$ and $De_m = 1/2$, 1, 2 and 4 and we perform the simulations with *M*=2, 4, 6, 8… until convergence of $\hat{G}_{[M]}$ in five significant digits is achieved. The results for $\tilde{G}_{[M]} = (1-\eta)\hat{G}_{[M]}/(2De_m)$ are reported in Table 1, where one can see that only a few spectral coefficients are adequate for convergence with the targeted accuracy. As expected, the increase of $De_m$ requires an increase of *M* to achieve the desired accuracy. It is interesting that the lowest-order approximation, $\tilde{G}_{[2]}$, gives the correct result within an absolute relative error of 1%, 0.4%, 1.2%, and 2% for $De_m = 1/2$, 1, 2, and 4, respectively, compared to the converged numerical value for G.

| M<br>$De_m$ | 2 | 4 | 6 | 8 | 10 | 12 | 14 |
|---|---|---|---|---|---|---|---|
| 1/2 | 1.0670 | 1.0782 | 1.0784 | 1.0784 | | | |
| 1 | 0.96562 | 0.96792 | 0.96929 | 0.96931 | 0.96931 | | |
| 2 | 0.91778 | 0.90338 | 0.90690 | 0.90721 | 0.90722 | 0.90722 | |
| 4 | 0.90354 | 0.88365 | 0.88419 | 0.88548 | 0.88570 | 0.88573 | 0.88573 |

**Table 1:** $\tilde{G}_{[M]} = -P'(Z)/H^3(Z)$ for $\eta=0.4$ ($c=2/3$) according to the fully spectral solution of Eqs. (5.9a,b)

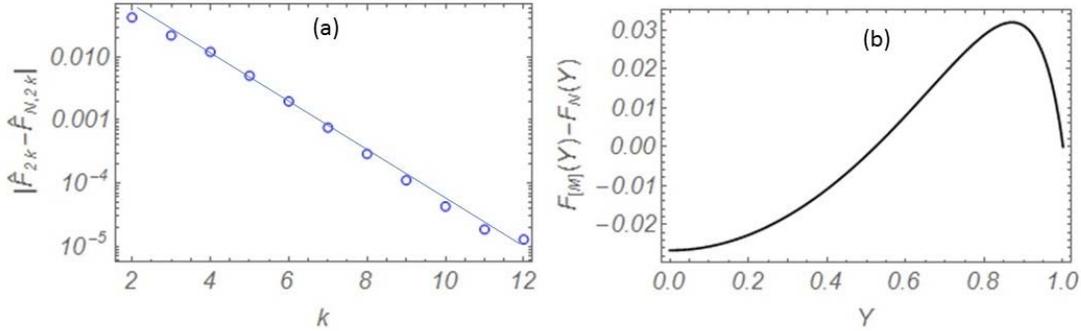

**Figure 3:** (a) The magnitude of the spectral – Legendre coefficients for $F_{[M]}(Y)$-$F_N(Y)$ with 12 spectral coefficients (M=12) (the straight line is a guide to the eye); (b) The difference of $F_{[M]}(Y)$ from the Newtonian value $F_N(Y)$ as function of the Y-coordinate. Parameters are $De_m=2$ and $\eta=0.4$.

The spectral accuracy of the code can be very easily confirmed by monitoring the magnitude of the spectral coefficients. For instance, in Figure 3a, the magnitude of the spectral coefficients of $\left| F_{[M]}(Y) - F_N(Y) \right|$, are shown in logarithmic scale for $De_m = 2$ and $c = 2/3$ ($\eta = 0.4$) using twelve Legendre modes (M=12). One can see the linear decrease of $\ln\left(| \hat{F}_{2k} - \hat{F}_{N,2k} |\right)$, i.e. the exponential decrease of $| \hat{F}_{2k} - \hat{F}_{N,2k} |$ and therefore the flow field is spectrally resolved with very high accuracy (note that $\hat{F}_{N,0} = 1 + De_m$, $\hat{F}_{N,2} = -De_m$, and $\hat{F}_{N,2k} = 0$ for $k > 1$). Also, the very small magnitude of the spectral coefficients shows that $F$ is slightly affected by the fluid viscoelasticity compared to the corresponding Newtonian solution. The difference $F_{[M]}(Y) - F_N(Y)$ is shown as function of Y in Figure 3b. The results reveal two regions; the first is located around the midplane of the channel, where the fluid decelerates, and the second is near the walls, where the fluid accelerates because of the mass conservation at each cross-section of the channel. Despite the small magnitude of $F(Y) - F_N(Y)$, its effect on the accurate prediction of the polymer extra-stresses near the wall is significant, as discussed below.

The results for the polymer extra-stresses are shown in Figure 4. The simulations to calculate $F$ are performed for $De_m$=1, 2, and 4, using $M$=10, 12, and 16, respectively, to achieve convergence in five significant digits; the polymer viscosity ratio is $\eta = 0.4$. From $F$, we recover $\tilde{\Psi}'(Y) = (F(Y)-1)/(2De_m)$ which we substitute in Eq. (5.4a-c) along with $\Lambda = 8$ to calculate $\tilde{\tau}_{ss}$, $\tilde{\tau}_{ns}$ and $\tilde{\tau}_{nn}$. The results for $\tilde{\tau}_{ss}/De$, $\tilde{\tau}_{ns}$ and $\tilde{\tau}_{nn}$ are illustrated in Figures 4a, 4b, and 4c, respectively, as functions of the distance from the midplane *Y*. It can be observed that all components of the modified extra-stress tensor are of similar magnitude. This reflects the successful scaling of the original components of the polymer extra-stress tensor $\tau_{zz}$, $\tau_{yz}$, and $\tau_{yy}$ with $H^4$, $H^2$, and $H^0$, respectively (see Eqs. (3.4) and (3.5)), and the significance of working with $\tau_{ss}$, $\tau_{ns}$ and $\tau_{nn}$ instead of $\tau_{zz}$, $\tau_{yz}$ and $\tau_{zz}$ in resolving accurately the polymer extra-stresses. Additionally, the stresses change only slightly with distance from the midplane, but exhibit significant variations as the wall is approached, indicating the presence of a boundary layer near the wall for all components. This boundary layer becomes thinner as the $De_m$ increases.

The monotonic variation of $\tilde{\tau}_{ss}$, $\tilde{\tau}_{ns}$ and $\tilde{\tau}_{nn}$ with respect to *Y* is also worth mentioning, indicating that

$$0 \leq \tilde{\tau}_{ss}(Y) \leq 2De\tilde{\Gamma}^2, \quad -\tilde{\Gamma} \leq \tilde{\tau}_{ns}(Y) \leq \tilde{\Gamma}, \quad \tilde{\tau}_{nn}(0) \leq \tilde{\tau}_{nn}(Y) \leq 0 \quad (5.12)$$

where $\tilde{\Gamma} \equiv \tilde{\Psi}''(-1) = -\tilde{\Psi}''(1) > 0$; recall that $\tilde{\tau}_{ss}(0) = \tilde{\tau}_{ns}(0) = \tilde{\tau}_{nn}(\pm 1) = 0$, $\tilde{\tau}_{ss}(\pm 1) = 2De\tilde{\Gamma}^2$, and $\tilde{\tau}_{ns}(\mp 1) = \pm \tilde{\Gamma}$. In other words, the results for the modified components of the polymer extra-stresses are always bounded from below by their corresponding values at the midplane, whereas they are bounded from above by their values at the walls. Notably, these quantities are independent of the shape function, which is responsible for the significant increase in both viscous and viscoelastic stresses. Indeed, the original shear stress at the wall is $H^{-2}(Z)\partial^2\Psi/\partial^2Y\big|_{Y=1}$ which according to the new exact solution varies from $\tilde{\Psi}''(Y)$ at *Z*=0 to $\Lambda^2\tilde{\Psi}''(Y)$ at *Z*=1. We emphasize that Eq. (5.12) is based on observation from the numerical results calculated here based on the exact solution, as well as the results calculated using the pseudospectral code for the solution of the full lubrication equations (Eqs. (3.6)-(3.10)); no theoretical proof is provided.

Furthermore, Figure 4 illustrates that the magnitude of the polymer extra-stresses at the walls increases with increasing $De_m$, deviating progressively from the corresponding values predicted by approximating the viscoelastic velocity profile with the Newtonian velocity profile. Specifically, using $\tilde{\Psi}(Y) \approx \Psi_N(Y)$ in Eq. (5.4a-c), we find $\tilde{\tau}_{ss}(1)/De = 9/2$, $\tilde{\tau}_{ns}(1) = -3/2$ and $\tilde{\tau}_{nn}(1) = 0$; recall that $\tilde{\Psi}(Y) = \Psi_N(Y)$ is the exact solution for the Newtonian fluid with $\Lambda \geq 1$, for the Oldroyd-B model with $\Lambda = 1, 0 < \eta < 1$, and for the Oldroyd-B model with $\Lambda > 1, \eta = 0$. Comparing these values with the actual values at the wall, it becomes evident that the approximation $\tilde{\Psi}(Y) \approx \Psi_N(Y)$ is highly inaccurate for evaluating the extra-stress at the walls and within the boundary layer. The approximation $\tilde{\Psi}(Y) = \Psi_N(Y)$ is valid only for the uncoupled problem, i.e., for $De \geq 0, \eta = 0$, where no interaction exists between the momentum balance and the polymer extra-stresses induced by the flow deformation.

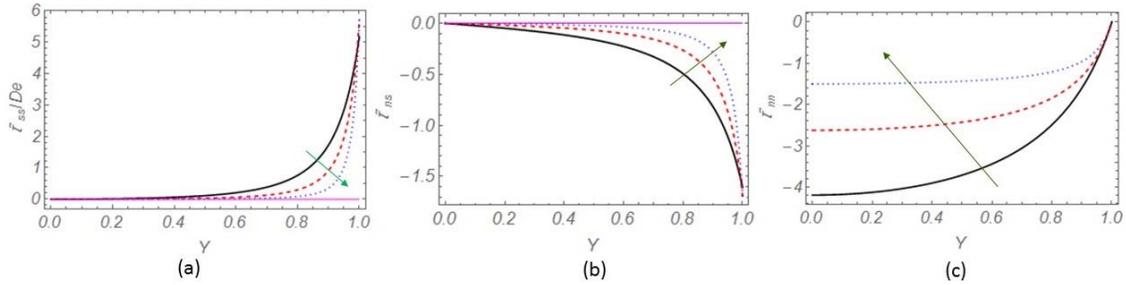

Figure 4: Modified polymer extra-stress components according to the exact solution, Eq. (5.4a-c), as functions of Y, for $De_m=1$ (solid black lines), $De_m=2$ (dashed red lines) and $De_m=4$ (dotted blue lines); parameters are $\Lambda=8$ and $\eta=0.4$. The arrow shows in the direction of increasing $De_m$.
(a) $\tilde{\tau}_{ss}/De$; (b) $\tilde{\tau}_{ns}$; (c) $\tilde{\tau}_{nn}$.

## 6. Results: pressure gradient and average pressure drop

As mentioned above, the average pressure drop, $\Delta\Pi$, in the hyperbolic channel is one of the most interesting and important features for this type of flow. Since the main novelty of this work is the exact analytical solution for the polymer extra-stresses, we focus on the reduced average pressure-drop, $\Delta\Pi/\Delta\Pi_N$, derived based on the exact solution. To determine the range of validity of the solution in terms of the dimensionless numbers and parameters appearing into the equations, we compare the new results with:

(i) those derived analytically, using the high-order asymptotic solution derived in §5.1, and

(ii) those calculated numerically using the pseudospectral code for the solution of the new set of lubrication equations, described in §5.2.

All theoretically equivalent formulations derived in Section 4 are utilized below to demonstrate the consistency of the theoretical analyses and the correctness of the methods of solution and the results. Specifically, Eqs. (4.1), (4.5), and (4.6) as derived from the momentum balance, total force balance, and mechanical energy of the flow, respectively, are used to calculate $\Delta\Pi / \Delta\Pi_N$ where $\Delta\Pi_N$ is given by Eq. (4.2).

First, we start using the high-order perturbation solution found in subsection 5.1 to determine the reduced pressure drop in series form. All formulations give the same result:

$$\frac{\Delta\Pi}{\Delta\Pi_N} \approx 1 - 3\eta De_m + \frac{216}{35}\eta De_m^2 + \frac{36}{35}(\eta-11)\eta De_m^3 + \left(\frac{216}{11} - \frac{90477}{13475}\eta + \frac{648}{385}\eta^2\right)\eta De_m^4$$
$$+\eta De_m^5\left(\pi_5(\eta) + \pi_6(\eta)De_m + \pi_7(\eta)De_m^2 + \pi_8(\eta)De_m^3\right)$$
(6.1)

where $\pi_k = \pi_k(\eta)$, $k=5,6,7$ and 8 are provided in Appendix C. Eq. (6.1) up to $O(De_m^8)$ fully agrees with the corresponding equation derived by Housiadas & Beris (2024a,b,c), and with the equation derived by Boyko & Stone (2022) up to $O(De_m^3)$. Note that Housiadas & Beris (2024a) used a modified Deborah number equal to $3(\Lambda-1)De/4$, while Boyko & Stone (2022) used a Deborah number equal to $\Lambda De/2$. An interesting feature of Eq. (6.1) is that the dependence of the reduced pressure drop on the contraction ratio is hidden into the definition of the modified Deborah number, i.e. $\Delta\Pi / \Delta\Pi_N$ is a function of $De_m$ and $\eta$. Housiadas & Beris (2023; 2024a) also demonstrated that the accuracy of the asymptotic series Eq. (6.1) and its range of convergence increase as more terms in the series are taken into account. Furthermore, when Eq. (6.1) is processed with the Padé [N/N] diagonal approximant, a clear convergence is achieved even for large values of $De_m$ when at least the first five terms in the series given in Eq. (6.1) are considered for the construction of the approximant (i.e., for N=2); note that the best agreement with the numerical results is achieved for N=4.

Additionally, we use our new pseudospectral code to perform the simulations, and we calculate $\Delta\Pi / \Delta\Pi_N$ based on Eqs. (4.1), (4.5) and (4.6). The simulations are performed for $\eta$=0.4 and 0.5, $2 \leq \Lambda \leq 8$, and $0 \leq De \leq 0.5$. For higher values of the contraction ratio, the polymer viscosity ratio, or the Deborah number, either a lack of convergence of the fully implicit scheme with an absolute error of $10^{-12}$ occurs, or instabilities are observed; both scenarios are typically associated with the loss of positive definiteness of the conformation tensor. Simulation results that lack spectral accuracy (i.e., an exponential decrease in the magnitude of the first two-thirds of the Chebyshev coefficients for the streamfunction) are also considered unsuccessful. The main advantage of the new pseudospectral code, compared to that developed by Housiadas & Beris (2023), is its ability to resolve the modified polymer extra-stress components ($\tau_{ss}$, $\tau_{ns}$ and $\tau_{nn}$), which remain of order unity throughout the computational domain. For instance, it is significantly easier to numerically solve for $\tau_{ss}$ than for $\tau_{zz} = \tau_{ss}/H^4$; recall that $\tau_{ss}(Y, Z=0) = 9DeY^2/2$ (i.e., $0 \leq \tau_{ss}(Y, Z=0) \leq 9De/2$), and $0 \leq \tau_{ss}(Y,Z) \leq 2De\,\Gamma^2(Z)$ where $\Gamma(Z) \equiv \partial^2\Psi/\partial Z^2\big|_{Y=1}$.

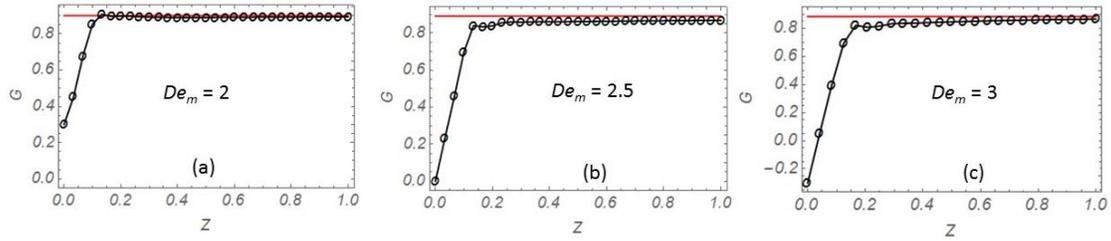

**Figure 5:** The modified pressure gradient as a function of the distance form the inlet
(a) $De_m$=2; (b) $De_m$=2.5; (c) $De_m$=3; parameters are $\eta$=0.4 and $\Lambda$=8.
Solid lines with dots: numerical pseudospectral results of the full lubrication equations
Straight (red) lines: based on the exact solution.
In the straight entry region $G$=3/2, while at the entrance of the hyperbolic section ($Z$=0)
a jump discontinuity occurs $G$ = (3/2)(1 − $\eta$ $De_m$).

The lack of convergence, the instabilities, or non-spectral accuracy observed at higher values of Λ, *De*, or *η* are related to the adverse pressure gradient that occurs when $1 - \eta(\Lambda - 1)De < 0$ (see Eq. (3.30)); for instance for $\eta = 0.4$, an adverse pressure gradient appears for $De_m > 2.5$. In Figure 5, we present the simulation results for *G* as a function of *Z* for $De_m = 2$, 2.5, and 3.0, using $\eta = 0.4$ and $\Lambda = 8$. The corresponding values from the exact solution are also shown in the figure as straight lines (recall that, according to the new exact solution, the flow variables are independent of *Z*). The chosen values for the modified Deborah number correspond to a jump of the pressure gradient at the inlet which is favorable to the flow (for $De_m = 2$), neutral (for $De_m = 2.5$) and adverse (for $De_m = 3$). Running the simulations, we observed that the convergence of the results becomes progressively more difficult as we exceed the threshold $De_m = 1/\eta = 2.5$ and as we continue increasing the modified Deborah number beyond this threshold. In all cases, two major observations are made: First, the flow always establishes a linear increase in the modified pressure gradient *G* with a large slope as the distance from the inlet increases. Second, the modified pressure gradient approaches the corresponding value predicted by the exact solution, $\tilde{G}$. Note, however, that the influence of the recovering region—namely, the region where $G$ increases with Z—on the reduced average pressure drop $\Delta\Pi/\Delta\Pi_N$ is negligible because $G$ must be divided by $H^3$ before being integrated with respect to Z to give the final result for $\Delta\Pi$ (see Eq. (4.1)).

Using the exact solution for the extra-stresses in Eqs. (4.1), (4.5), and (4.6), yields:

$$\frac{\Delta\Pi}{\Delta\Pi_N} = \frac{2}{3}\tilde{G} \tag{6.2}$$

$$\frac{\Delta\Pi}{\Delta\Pi_N} = -\frac{2}{3}\tilde{\Psi}''(1) - 2\eta(\Lambda-1)\int_0^1 \tilde{\tau}_{ss}(Y)dY \tag{6.3}$$

$$\frac{\Delta\Pi}{\Delta\Pi_N} = \frac{4}{3}(1-\eta)\int_0^1 \left(\tilde{\Psi}''(Y)\right)^2 dY + \frac{2\eta}{3De}\int_0^1 \tilde{\tau}_{ss}(Y)dY - \frac{8\eta}{3}(\Lambda-1)\int_0^1 \tilde{\Psi}'(Y)\tilde{\tau}_{ss}(Y)dY \tag{6.4}$$

The first term on the right-hand-side of Eq. (6.3) corresponds to the viscous forces exerting from the fluid to the wall(s) of the channel, while the second term corresponds to the bulk contribution due to the viscoelastic of the fluid. Notably, the viscous term is positive, whereas the viscoelastic term is negative (recall that $\tilde{\tau}_{ss}(Y) \geq 0$ for $|Y| \leq 1$). Similarly, the first term on the right-hand side of Eq. (6.4) represents the viscous dissipation, the second corresponds to the elastic dissipation, and the third denotes the work done by the elastic forces. As shown in Eq. (6.4), the viscous and elastic contributions are positive, while the work done by the elastic forces is negative for any $\Lambda > 1$.

With a goal to achieve further analytical progress, we can use $\tilde{G} \approx \tilde{G}_{[2]}$ where $\tilde{G}_{[2]}$ is given by Eq. (5.11). Thus, Eq. (6.2) gives:

$$\frac{\Delta\Pi}{\Delta\Pi_N} \approx 1 - \eta + \frac{\eta}{1 + 3De_m + 126De_m^2/35 + 54De_m^3/35} \tag{6.5}$$

Moreover, as a first approximation we can assume that $\tilde{\Psi}(Y) \approx \Psi_N(Y)$. However, as clearly demonstrated in Section 5.3, this approximation is very poor and cannot be used for accurately evaluating the polymer extra-stresses in the coupled case (i.e., for $0 < \eta \leq 1$). Thus, while Eq. (6.3) may not yield accurate results (mainly because $|\tilde{\Psi}''(1)| \gg |\Psi_N''(1)| = 3/2$), Eq. (6.4) involves the polymer extra-stresses and the pure viscous stresses in a non-linear way, which can lead to cancellation errors. Substituting $\tilde{\Psi}(Y) \approx \Psi_N(Y)$ in Eqs. (6.3) and (6.4) and performing the integrations, we obtain the following results, respectively:

$$\frac{\Delta \Pi}{\Delta \Pi_N} \approx 1 + \eta \left( \frac{\sqrt{3} \coth^{-1}\left[\sqrt{1 + \frac{2}{3De_m}}\right]}{De_m^{1/2}(2+3De_m)^{3/2}} - \frac{3(1+3De_m)}{2(2+3De_m)} \right) \qquad (6.6)$$

$$\frac{\Delta \Pi}{\Delta \Pi_N} \approx 1 + \eta \left( \frac{(5+12De_m)\tanh^{-1}\left[\sqrt{\frac{3De_m}{2+3De_m}}\right]}{\sqrt{3}\left(De_m(2+3De_m)\right)^{3/2}} - \frac{5+De_m(7+6De_m)}{2De_m(2+3De_m)} \right) \qquad (6.7)$$

To assess the accuracy of Eqs. (6.5)-(6.7), we present results for $\Delta \Pi / \Delta \Pi_N$ as function of $De_m$ using the following methods:

(i) the 8[th]-order perturbation solution (Eq. (6.1)), post-processed with the Padé [4/4] approximant (black solid lines),

(ii) the pseudospectral simulations of the full lubrication equations (Eqs. (3.6)-(3.9)) (star symbols),

(iii) the new exact solution along with the fully spectral simulations for the numerical evaluation of the streamfunction and the pressure gradient (open symbols), such as the results given in Table 1, and

(iv) Eq. (6.7) which has been derived from the exact new solution, the mechanical energy decomposition, and by approximating the streamfunction with the corresponding streamfunction for a Newtonian fluid.

Specifically, in Figure 6a, we present the reduced average pressure drop for $\eta = 0.5$ and $\Lambda = 4.5$ in the range $0 \leq De_m \leq 1.5$; these values are based on the experiments conducted in the 3D axisymmetric geometry by James & Roos (2021) (see subsection 7.3 for more details). In Figure 6b, we present the results for $\eta = 0.4$ and $\Lambda = 8$ in the range $0 \leq De_m \leq 4$. In all cases, the results show the decrease of $\Delta \Pi / \Delta \Pi_N$ with increasing $De_m$ (Boyko & Stone 2022; Housiadas & Beris 2023, 2024a,d). If Figure 6a, the excellent agreement of all methods is evident, as the differences between the results are hardly discernible. Note that the range of parameters shown in Figure 6a is within the region $1 - \eta De_m > 0$, in which the jump discontinuity in the slope of the geometry at the inlet does not does not induce an adverse pressure gradient in the flow. Similarly, in Figure 6b we observe that when the dimensionless parameters are within the region $1-\eta De_m > 0$, the results obtained with the four methods mentioned above are in remarkable agreement. However, for $1-\eta De_m < 0$, the results from the pseudospectral simulations (star symbols) using the full set of lubrication equations start to deviate from those obtained by the other methods.

Note also that among the approximate analytical formulas Eqs. (6.5)-(6.7), Eq. (6.6) is the least accurate, while Eq. (6.7) is the most accurate when compared to the converged results obtained from the pseudospectral simulations. However, all formulas deviate progressively from the exact simulation results as $De_m$ increases, which is to be expected due to the failure of the Newtonian velocity profile to accurately capture the large velocity gradient at the channel walls. This gradient becomes larger as $De_m$ increases. Thus, we conclude this section by reiterating that Eq. (6.7) can be reliably used to calculate the reduced average pressure drop with high accuracy within the parameter range $1-\eta De_m \geq 0$, and with acceptable accuracy for values somewhat above the threshold $1 = \eta De_m$ (as seen in Figure 6b).

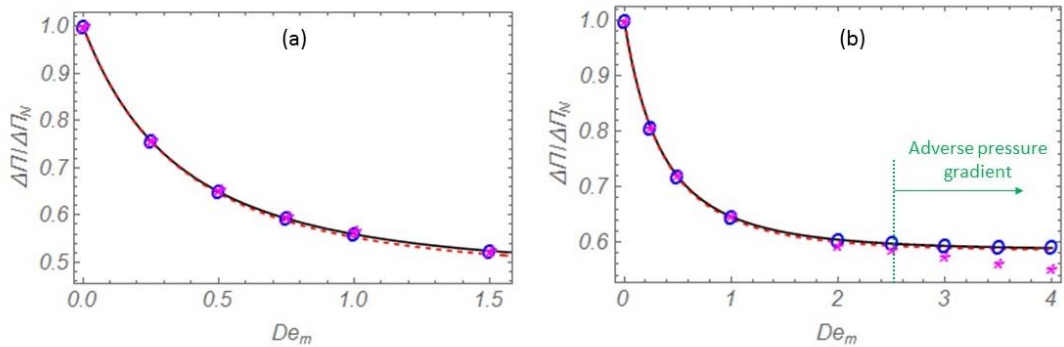

**Figure 6:** Reduced pressure drop vs the modified Deborah number;
(a) $\eta$=0.5 and $\Lambda$=4.5 (parameters based on the experiments by James & Roos (2021));
(b) $\eta$=0.4 and $\Lambda$=8.
Open symbols: new exact solution,
Star symbols: full pseudospectral simulations,
Black line: 8$^{th}$ order asymptotics with Padé [4/4] approximant;
Red line: Eq. (6.7) based on the new exact solution and the mechanical energy;
In Figure 6b, the region above which an adverse pressure gradient appears, is marked with a green vertical line.

## 7. Discussion, limitations, and properties of the exact solution

Starting from the original lubrication equations governing the steady, isothermal, and incompressible flow of an Oldroyd-B fluid in a planar 2D configuration, we derived a new set of lubrication equations in terms of new spatial coordinates that map the varying shape of the channel's fixed walls into a constant configuration, the streamfunction, as well as new components of the polymer extra-stress tensor which are a linear combination of the components of the original polymer extra-stress tensor. The new set of governing equations can be easily shown to be identical to the set of lubrication equations derived independently by Hinch, Boyko & Stone (2024) using different methods and techniques (see Appendix A). For the particular case of a hyperbolic geometry, the ratio of the channel height at the inlet to the channel height at the outlet, Λ, serves as the main geometrical parameter distinguishing a straight channel (Λ=1) from an expanding channel (0<Λ<1) and a contracting one (Λ>1). The latter configuration has been widely employed in experimental studies, practical applications, and theoretical analyses, and is the focus of the current work.

For Λ=1, and ignoring the derivatives of the field variables along the main flow direction, $Z$, except for the pressure gradient—which in no circumstances can be eliminated from the equations—one derives the classic Poiseuille solution for the Oldroyd-B fluid in a straight channel. Since the Z-derivatives of the velocity components are neglected in the momentum balance, this solution does not account for any entry or exit effects in the channel. Furthermore, eliminating the Z-derivatives from the constitutive model precludes the imposition of any type of boundary conditions on the components of the polymer extra-stress tensor.

The idea of ignoring the Z-dependence on the flow variables leads to the classic Poiseuille solution, which as commonly described in the literature, corresponds to a "fully developed flow", meaning that the flow does not vary downstream. This solution is exact—it precisely satisfies the mass and momentum balances as well as the constitutive model for any finite value of the Deborah number—but its range of validity is limited. This limitation stems from viscoelastic instabilities that arise at high Deborah numbers, even for slow flows (i.e., under creeping conditions)—in same cases viscoelastic turbulence can be even induced. To date, the exact range of validity of the classic Poiseuille solution has not been established theoretically.

For Λ>1, it can be argued that the flow can never be fully developed, as the variation in geometry along the Z-direction continuously alters the field variables (velocity, pressure, and polymer extra-stresses) downstream. However, the new set of lubrication equations, optimized to account for the hyperbolic properties of the geometry, admits an exact solution in a manner analogous to that of the straight channel—by neglecting the Z-derivatives of all dependent variables. This solution identically satisfies the non-trivial components of the Oldroyd-B model and is theoretically valid for any value of the Deborah number. Hence, it become clear that the independence of $\Psi, \tau_{ss}, \tau_{ns}$ and $\tau_{nn}$ from the mapped Z-coordinate is fundamentally different from the independence of $U, V, \tau_{zz}, \tau_{yz}$ and $\tau_{yy}$ from the original z-coordinate. In other words, the original flow variables change with distance from the inlet in a manner that is concealed within the new variables but can be uncovered using Eqs. (3.1) and (3.2) for the original velocity components, and Eq. (3.5) for the original components of the polymer extra-stress tensor:

$$U(Y,Z) \approx \frac{\tilde{\Psi}'(Y)}{H(Z)}, \quad V(Y,Z) \approx -(\Lambda-1)Y H(Z)\tilde{\Psi}'(Y), \quad \tau_{zz}(Y,Z) \approx \frac{\tilde{\tau}_{ss}(Y)}{H^4(Z)},$$

$$\tau_{yz}(Y,Z) \approx \frac{\tilde{\tau}_{ns}(Y)-(\Lambda-1)Y\tilde{\tau}_{ss}(Y)}{H^2(Z)}, \quad \tau_{yy}(Y) \approx \tilde{\tau}_{nn}(Y) - 2(\Lambda-1)Y\tilde{\tau}_{ns}(Y) + (\Lambda-1)^2 Y^2 \tilde{\tau}_{ss}(Y)$$

(7.1)

where the spatial dependences of all field variables are given explicitly for clarity. We emphasize that precisely as for the straight channel case, the exact range of applicability of the analytical solution in terms of the Deborah number, cannot be determined analytically due to the appearance of viscoelastic instabilities.

This situation is also analogous to the simple Poiseuille solution for a Newtonian fluid in a straight 2D channel, i.e., Eq. (3.13), or in a straight axisymmetric cylindrical pipe. One can easily verify that Eq. (3.13) precisely satisfies not only the classic lubrication equations but also the full Navier-Stokes equations for arbitrary Reynolds numbers. However, it is well-known that as the Reynolds number increases, the flow becomes unstable; eventually, at sufficiently large Reynolds numbers, turbulence develops, requiring a substantial increase in pressure drop to maintain a constant flow rate through the channel. Note that the exact range of validity of the classic Poiseuille solution has not been determined theoretically although experimentally the laminar, intermittent and turbulent regions in terms of the Reynolds are well-known.

### 7.1. Limiting cases of the exact solution

The properties of the exact solution, i.e., Eq. (5.4a-c), are investigated here to confirm the consistency of the solution in special cases and identity its weak features. As mentioned in Subsection 5.1, when $De = 0$ the solution reduces to the Newtonian solution in a hyperbolic channel, given by Eqs. (3.13) and (3.14), which are valid for $\Lambda > 1$. For a straight channel $\Lambda = 1$ and $De > 0$, the solution simplifies to Eqs (3.16)-(3.17), as it should.

Furthermore, when Eq. (5.4a–c), the corresponding momentum balance (Eq. (5.7)), and the equation for the modified pressure gradient (Eq. (5.8)) are expanded in terms of the original Deborah number, the high-order asymptotic solution derived in Subsection 5.1 is recovered. Hence, the exact solution is more general than that derived for weakly viscoelastic fluids and is expected to remain valid and accurate over an extended range of Deborah numbers compared to the asymptotic solution. Additionally, since $\tilde{\Psi}'(Y=1)=0$, it is straightforward to confirm that Eq. (5.4a-c) reduces to the general exact solution of the Oldroyd-B model along the walls, expressed in terms the shear stress at the wall (i.e., Eq. (3.27)).

However, it appears that Eq. (5.4a–c) does not satisfy the boundary conditions for the polymer extra-stresses at the inlet of the hyperbolic section of the channel (*Z*=0), as given by Eq. (3.17). This is not surprising, as neglecting the Z-derivatives of all field variables makes it impossible to satisfy any boundary conditions. Consequently, entry effects and the development of the extra-stresses with distance from the inlet cannot be accounted for. In other words, the solution presented here is a particular solution of the full differential equations describing the Oldroyd-B model, i.e., Eqs. (3.7)–(3.9), and cannot, under any circumstances, be considered the general solution to which specified initial conditions can be applied. Worth noting is that due to the hyperbolic nature of the constitutive model, imposing the initial conditions given by Eq. (3.17) on $\tau_{ss}$, $\tau_{ns}$ and $\tau_{nn}$ is equivalent to ensuring the continuity of polymer extra-stresses between the entrance region and the hyperbolic region, i.e., along the Z-direction.

To assess the range of validity of Eq. (5.4a–c), one must compare it with the full analytical solution of Eqs. (3.6)–(3.10), which is, of course, not known, or alternatively with results obtained from numerical simulations, as performed in Section 6 for the average pressure drop in the channel. Additional insight can also be gained by following the analysis presented for the evolution of polymer-extra stresses along the symmetry plane of a planar 2D channel by Housiadas & Beris (2023, 2024a), and, similarly, along the symmetry axis of an axisymmetric 3D pipe by Housiadas & Beris (2024d) and Sialmas & Housiadas (2025). Specifically, at the midplane *(Y*=0), $\tilde{\Psi}''(0)=0$ due to the symmetry of the flow field and Eq. (5.4a-c) reduces to:

$$\tilde{\tau}_{ss}(0) = \tilde{\tau}_{ns}(0) = 0, \quad \tilde{\tau}_{nn}(0) = -\frac{2(\Lambda-1)\tilde{\Psi}'(0)}{1+2(\Lambda-1)De\,\tilde{\Psi}'(0)} \qquad (7.2\text{a-c})$$

Comparing Eqs (7.2a,b,c) with (3.22)-(3.24), one can see that the *ss*- and *ns*-components of the polymer extra-stress tensor at the midplane are the same with the exact analytical solution, but the *nn*-components do not entirely agree ($\tilde{\tau}_{nn}(0) \neq \tau_{nn}(0,Z)$) since the dependence on *Z* is clear. In order to address the how close is $\tilde{\tau}_{nn}(0)$ to the exact $\tau_{nn}(0,Z)$, it is necessary to determine the fluid velocity at the midplane, which, of course, requires solving all the governing equations. However, valuable information can be extracted from the uncoupled case, i.e., for $\eta=0$, where the velocity profile matches the Newtonian velocity profile, $\tilde{\Psi}(Y) = \Psi_N(Y)$. Substituting $\tilde{\Psi}'(0) = \Psi'_N(0) = 3/4$ and $u(Z) = \Psi'_N(0)/H(Z) = 3/(4H(Z))$ in Eqs. (5.4c) and (3.25c), we find, respectively:

$$\tilde{\tau}_{nn}(0) = -\frac{3(\Lambda-1)/2}{1+3De_m/2} \tag{7.3}$$

$$\frac{\tau_{nn}(0,Z)}{\tilde{\tau}_{nn}(0)} = 1 - H^a(Z), \quad a = 2 + \frac{4}{3De_m} \tag{7.4}$$

where the limits $\lim_{De_m \to 0^+} a = \infty$ and $\lim_{De_m \to \infty} a = 2$ are worth noting. From Eq. (7.4), it is not difficult to confirm that:

$$\lim_{De_m \to 0^+} \frac{\tau_{nn}(0,Z)}{\tilde{\tau}_{nn}(0)} = \begin{cases} 0, & Z=0 \\ 1, & 0 < Z \leq 1 \end{cases} \tag{7.5}$$

and

$$\lim_{De_m \to \infty} \frac{\tau_{nn}(0,Z)}{\tilde{\tau}_{nn}(0)} = 1 - H^2(Z), \quad 0 \leq Z \leq 1 \tag{7.6}$$

Notice that the quantity $1-H^2(Z)$ deviates very little from unity; for instance for $\Lambda = 8$ and $\Lambda = 10$, it decreases monotonically from 1 to 0.984375 and from 1 to 0.99, respectively. Hence, $\tau_{nn}(0,Z)$ and $\tilde{\tau}_{nn}(0)$ are asymptotically equivalent quantities ($\tau_{nn}(0,Z) \sim \tilde{\tau}_{nn}(0)$) for both $De_m \ll 1$ and $De_m \gg 1$ everywhere in the whole domain, with the only exception being the inlet of the hyperbolic section of the channel ($Z = 0$) and for a negligible $De_m$.

Results for $\tilde{\tau}_{nn}(0)$ and $\tau_{nn}(0,Z)$ are shown in Figure 7a, along with the corresponding results for $\tilde{c}_{nn}(0)$ and $\tilde{c}_{nn}(0,Z)$ in Figure 7b, using a contraction ratio $\Lambda = 10$ and $De_m \equiv (\Lambda-1)De$ = 1/2, 1, 2, and 10; these values correspond to *De*=1/18, 1/9, 2/9 and 10/9 illustrating that a large value of the modified Deborah number does not necessarily mean that the original Deborah number is large too. In all cases, one observes a boundary layer near *Z*=0 where $\tau_{nn}(0,Z)$ decays rapidly with increasing *Z*, eventually approaching $\tilde{\tau}_{nn}(0)$. Additionally, it is observed that the smaller the modified Deborah number, the more rapid the decay of the initial condition. Furthermore, for large values of $De_m$, the differences between $\tilde{\tau}_{nn}(0)$ and $\tau_{nn}(0,Z)$ are negligible. Finally, we mention that the variation and the magnitude of $\tau_{nn}(Y=0,Z)$ and $\tilde{\tau}_{nn}(Y=0)$ have no influence on the corresponding values for $\tau_{ss}(Y=0,Z) = \tilde{\tau}_{ss}(Y=0) = 0$ and $\tau_{ns}(Y=0,Z) = \tilde{\tau}_{ns}(Y=0) = 0$.

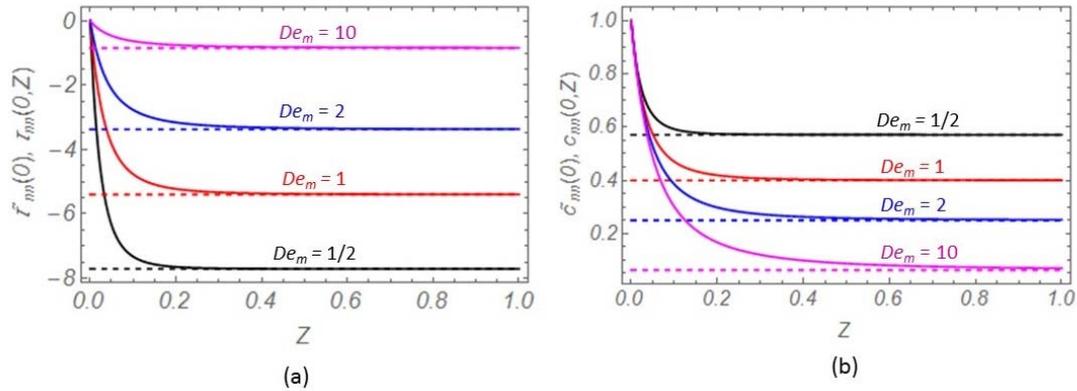

Figure 7: (a) The exact (solid lines) and approximate (dashed lines) values of the *nn*-component of the modified polymer extra-stress tensor at the midplane of the hyperbolic channel, $\tau_{nn}(Y=0,Z)$ and $\tilde{\tau}_{nn}(Y=0)$, respectively, for Λ=10 and η=0 (the latter implies a Newtonian velocity profile; see Eq. (3.13)). (b) The same results as in (a) shown for the corresponding components of the conformation tensor.

We emphasize that the analysis presented above strictly applies to the uncoupled case, i.e., for *η*=0, where Newtonian kinematics are imposed in the constitutive model to calculate the polymer molecules' response to flow deformation. For *η* > 0, the coupling between the polymer extra-stress, viscous stresses, and the pressure gradient through the momentum balance becomes strongly non-linear, even under the lubrication approximation. In this case, as demonstrated in subsection 5.3, when the Deborah number increases, the Newtonian velocity profile fails to accurately describe the velocity gradients and the induced extra-stress near the walls.

It is worth mentioning that $\tau_{nn}$ does not enter directly the mass and momentum balances that govern the flow, as Eqs. (2.8)-(2.10), or the equivalent Eq. (3.6), show. Consequently, the general formulas for the average pressure drop derived in Section 4 do not involve $\tau_{nn}$ whatsoever. Thus, the error introduced due to the approximation of $\tau_{nn}(Y,Z)$ with $\tilde{\tau}_{nn}(Y)$ has a minimal effect on the fluid velocity in the channel, and as shown above with the aid of the high-order asymptotic solution in terms of the Deborah number, for weakly viscoelastic fluids this approximation has no effect on the results. It also became clear above that the difference between $\tau_{nn}(Y,Z)$ with $\tilde{\tau}_{nn}(Y)$ was caused due to effect of the boundary condition at the inlet of the hyperbolic section of the channel; thus when this effect attenuates $\tau_{nn}(Y,Z)$ approaches $\tilde{\tau}_{nn}(Y)$.

## 7.2. On the dimensionless numbers and parameters

First, we need to summarize the conditions under which the classic lubrication equations for an Oldroyd-B fluid, Eqs. (2.8)-(2.13), or Eqs. (3.6)-(3.9), are valid. By construction the equations are valid for $\text{Re} \ll 1$, $\varepsilon \ll 1$ and $Wi \gg 1$. In other words, the equations are valid for slow flows where inertia can be neglected, long channels where entrance and exit effects can be ignored, and fluids exhibiting strongly non-linear viscoelasticity. Additionally, the contracting hyperbolic geometry allows the evaluation of the Hencky strain, $\varepsilon_H$, that can be achieved in experiments; for the planar 2D geometry the Hencky strain equals the logarithm of the contraction ratio, $\varepsilon_H = \ln(h_0^*/h_f^*) = \ln(\Lambda)$, while for the 3D axisymmetric geometry the corresponding Hencky strain is twice the logarithm of the contraction ratio, $\varepsilon_H = \ln((\pi R_0^{*2})/(\pi R_f^{*2})) = 2\ln(\Lambda)$, where $R_0^*$ and $R_f^*$ are the radius at the inlet and outlet cross sections of the pipe. Experimentally, much larger Hencky strains are attainted in 3D axisymmetric geometries that in planar geometries. Typical values for the contraction ratio in the 3D axisymmetric hyperbolic geometry is $3 \le \Lambda \le 7$ approximately; for instance, James & Roos (2021) and James & Tripathi (2023) used $\Lambda \approx 4.5$. However, large values of the contraction ratio are avoided in experiments because the stresses exerted by the fluid on the walls of the channel are so large that can cause fracture of the channel. For the same reason, low molecular weight fluids (i.e., fluids with relatively low viscosity) are used.

James & Roos (2021) in their experiments with dilute polymer solutions (Boger-type fluids), reported that the maximum attainable value of the Deborah number was $\lambda^* Q^* / (4 R_f^{*2} \ell^*) = 4.6$. When expressed in terms of the radius of the cross section at the inlet, $R_0^*$, the maximum attainable value of the Deborah number decreases to $De = \lambda^* Q^* / (2\pi R_0^{*2} \ell^*) \approx 0.144$ -- see Housiadas & Beris (2024d) and Sialmas & Housiadas (2025) for more details on the 3D axisymmetric geometry and the difference with the 2D planar case. These conditions clearly indicate that the maximum Deborah number, as defined here and in similar theoretical works in the literature, cannot exceed the value of 0.144 approximately. Similar conclusions hold for the 2D planar case.

We close this section by reiterating that the Deborah number is the ratio of the fluid's relaxation time to its characteristic residence time in the channel. Since the flow occurs at an almost zero Reynolds number, the characteristic residence time has a reference value at the midplane and increases as the fluid approaches the wall. This implies that the effective (or local) Deborah number decreases near the walls, approaching zero at the no-slip boundary. Consequently, a high-Deborah asymptotic analysis (as previously performed by Hinch et al. 2024) cannot be valid across the entire flow domain and completely fails near the walls—precisely in the region responsible for the most significant forces exerted by the fluid on the wall. These forces must be overcome by the imposed average pressure drop required to drive the flow.

From a rheological perspective the conditions $\mathrm{Re} \ll 1$, $\varepsilon \ll 1$, $Wi \gg 1$, and $De \gg 1$ in a pressure-driven confined flow clearly correspond to the flow of an elastic solid. This flow would be impossible to occur in a hyperbolic channel, for the reasons that have been identified by experimentalists and discussed above. In the context of a generalized Pipkin diagram, these conditions correspond to the upper-right part of the diagram for which the Oldroyd-B model (and similar models such as the Giesekus, FENE-P and FENE-CR, and Phan-Thien & Tanner (PTT) models) are not valid.

## 8. Conclusions

In a manner similar to the derivation of the classic Poiseuille solution for the steady incompressible flow of an Oldroyd-B fluid in a straight 2D symmetric channel, a new exact self-similar solution for the polymer extra-stress tensor in terms of the streamfunction was derived for the hyperbolic geometry. As with the solution for the straight channel which does not satisfy any boundary conditions (due to the elimination of the derivatives of the field variables with respect to the main flow direction), the same limitation applies here. Consequently, the derived similarity solution represents only a particular solution to the full lubrication equations. It is also worth noting that this solution cannot be derived using a regular perturbation scheme and is more general than the solution derived based on asymptotic techniques.

Although the new exact solution for the polymer extra-stresses appears valid for arbitrary (but finite) Deborah numbers, it is not expected to hold at sufficiently large De values (as confirmed here by the numerical simulations too). Apart from the failure to satisfy boundary conditions at the inlet, other reasons may include the loss of existence of solutions of the lubrication equations, the general invalidity and limitations of the lubrication equations, the emergence of instabilities as De increases (which may even lead to turbulence), and other factors.

Focusing on the reduced average pressure drop, $\Delta\Pi/\Delta\Pi_N$, required to maintain a constant flow rate through the channel, we exploited the exact solution and derived a variety of approximate analytical formulas for $\Delta\Pi/\Delta\Pi_N$, as functions of $De_m$ and $\eta$. The most accurate formula was based on the mechanical energy decomposition of the flow, Eq. (6.7), achieving excellent agreement with the pseudospectral results within the parameters range $1-\eta(\Lambda-1)De > 0$. Notably, convergent numerical results, based on spectral criteria, were obtained only slightly above the threshold $1 = \eta(\Lambda-1)De$ beyond which an adverse pressure gradient appears due to the discontinuity in the slope of the shape function at the inlet of the hyperbolic section of the channel.

Exact analytical solutions of viscoelastic flows in confined non-uniform geometries are not available in the literature of non-Newtonian fluid mechanics (at least as far as the author of this work is aware of). Thus, the exact analytical solution found here provides a unique opportunity for advancements and further theoretical developments in viscoelastic flows within hyperbolic geometries. As such, the solution is of great importance for understanding viscoelasticity in internal flows, developing new analytical solutions in similar geometries, developing and validating new numerical methods, assessing the performance of approximate analytical methods (such as asymptotic methods), conducting stability analysis of the flow, as well as for the study of heat and/or mass transfer in hyperbolic channels.

**Declaration of Interests**

The author reports no conflict of interest.

**Appendix A: Lubrication equations in term of the confirmation tensor**

Omitting the continuity equation, Eq. (2.8), and taking into account that $\partial P / \partial y = 0$, the original lubrication equations, Eqs (2.9) and (2.11)-(2.13), in terms of the components of the conformation tensor are:

$$-\frac{dP}{dz} + (1-\eta)\frac{\partial^2 U}{\partial y^2} + \frac{\eta}{De}\left(\frac{\partial c_{zz}}{\partial z} + \frac{\partial c_{yz}}{\partial y}\right) = 0, \quad (A.1)$$

$$c_{zz} + De\left(\frac{Dc_{zz}}{Dt} - 2c_{zz}\frac{\partial U}{\partial z} - 2c_{yz}\frac{\partial U}{\partial y}\right) = 0, \quad (A.2)$$

$$c_{yz} + De\left(\frac{Dc_{yz}}{Dt} - c_{yy}\frac{\partial U}{\partial y} - c_{zz}\frac{\partial V}{\partial z}\right) = 0, \quad (A.3)$$

$$c_{yy} + De\left(\frac{Dc_{yy}}{Dt} - 2c_{yz}\frac{\partial V}{\partial z} - 2c_{yy}\frac{\partial V}{\partial y}\right) = 1, \quad (A.4)$$

where the equations are valid over the domain $\{(y,z) \| -H(z) < y < H(z), 0 < z < 1\}$ and the material derivative is given below Eq. (2.13). Similarly, the new set of lubrications equations in terms of the components of the modified conformation tensor, the streamfunction, and the mapped coordinates, valid over the domain $\{(Y,Z), -1 < Y < 1, 0 < Z < 1\}$, are:

$$-\frac{dP}{dZ}H^3 + (1-\eta)\frac{\partial^3 \Psi}{\partial Y^3} + \frac{\eta}{De}\left(3(\Lambda-1)c_{ss} + \frac{\partial c_{ns}}{\partial Y} + \frac{1}{H}\frac{\partial c_{ss}}{\partial Z}\right) = 0 \quad (A.5)$$

$$c_{ss} + De\left(\frac{Dc_{ss}}{Dt} + 2(\Lambda-1)c_{ss}\frac{\partial \Psi}{\partial Y} - 2c_{ns}\frac{\partial^2 \Psi}{\partial Y^2} - 2\frac{c_{ss}}{H}\frac{\partial^2 \Psi}{\partial Z \partial Y}\right) = 0 \quad (A.6)$$

$$c_{ns} + De\left(\frac{Dc_{ns}}{Dt} + 2(\Lambda-1)c_{ns}\frac{\partial \Psi}{\partial Y} - c_{nn}\frac{\partial^2 \Psi}{\partial Y^2} + \frac{c_{ss}}{H}\left((\Lambda-1)\frac{\partial \Psi}{\partial Z} + \frac{1}{H}\frac{\partial^2 \Psi}{\partial Z^2}\right)\right) = 0 \quad (A.7)$$

$$c_{nn} + De\left(\frac{Dc_{nn}}{Dt} + 2(\Lambda-1)c_{nn}\frac{\partial \Psi}{\partial Y} + 2\frac{c_{nn}}{H}\frac{\partial^2 \Psi}{\partial Y \partial Z} + 2\frac{c_{ns}}{H}\left((\Lambda-1)\frac{\partial \Psi}{\partial Z} + \frac{1}{H}\frac{\partial^2 \Psi}{\partial Z^2}\right)\right) = 1 \quad (A.8)$$

where for any dependent flow variable $f = f(Y,Z)$ the material derivative is given in Eq. (3.3), or, alternatively, in its conservative form as follows:

$$\frac{Df}{Dt} = \frac{1}{H}\left(\frac{\partial}{\partial Z}\left(f\frac{\partial \Psi}{\partial Y}\right) - \frac{\partial}{\partial Y}\left(f\frac{\partial \Psi}{\partial Z}\right)\right) \quad (A.9)$$

The meaning of the underlying terms is the same to that in Eqs. (3.6)-(3.9).

Recalling that $(\Lambda - 1) = -H'(Z)/H^2(Z)$ and substituting:

$$\upsilon = -\frac{\partial \Psi}{\partial Z}, \; u = \frac{1}{H}\frac{\partial \Psi}{\partial Y}, \; A_{11} = \frac{c_{ss}}{H^4}, \; A_{12} = \frac{c_{ns}}{H^2}, \; A_{22} = c_{nn} \tag{A.10}$$

into Eqs. (A5)-(A9) yields the equations derived by Hinch, Boyko & Stones (2024) using orthogonal curvilinear coordinates (see Eqs. (2.1)-(2.4) in their paper). Consequently, the final equations developed here, despite employing different methods and techniques, are identical to those derived by Hinch, Boyko, & Stones (2024).

**Appendix B. Verification of the pseudospectral code for Eqs. (3.6)-(3.9)**

The correctness and spectral accuracy of the code developed in Subsection 5.2 is verified here by performing the simulations for four cases with increasing order of complexity. First, we note that for any known continuous function $f = f(Y)$ where $Y \in [-1,1]$, a truncated Chebyshev series expansion of $f$ is given by

$$f(Y) \approx f_{[M]}(Y) = \sum_{k=0}^{M} \hat{f}_k T_k(Y), \quad \hat{f}_k = \frac{2}{\pi c_k} \int_{-1}^{1} \frac{f(Y)T_k(Y)}{\sqrt{1-Y^2}} dY, \tag{B.1}$$

where in the expansion (or spectral) coefficients $\hat{f}_k$, $c_0 = 2$ and $c_k = 1$ for $k > 0$ [Orszag 1971; Hesthaven, Gottlieb, & Gottlieb, 2007]. In many cases, the spectral coefficients can be determined analytically by implementing the change of variable $Y = \cos(\theta)$; the latter simplifies the Chebyshev polynomials as $T_k(Y) = T_k(\cos(\theta)) = \cos(k\theta)$. Additionally, hereafter, we will use the term "with machine accuracy" to denote that the absolute error between a numerically calculated quantity and its exact formula is less than $10^{-15}$. Finally, we note that in the four cases described below, the full Newton scheme implemented at each grid point in the Z-direction, $Z_j$, finds the solution at the collocation points in the Y-direction in single iteration, while a second iteration confirms convergence with an accuracy $10^{-16}$.

*(i) Newtonian flow ($De = 0$) for any $\Lambda \geq 1$.*

In this case, the analytical solution is given by Eqs. (3.13)-(3.14). Application of (B.1) in the solution gives zero spectral coefficients except for $\hat{\Psi}_1 = 9/16$, $\hat{\Psi}_3 = -1/16$, $\hat{\tau}_{ns,1} = -3/2$, $\hat{\tau}_{nn,0} = -3(\Lambda-1)/4$ and $\hat{\tau}_{nn,2} = 3(\Lambda-1)/4$. Running the code for any value of the contraction ratio Λ, using five collocation points (M=4) in the Y-direction is sufficient to calculate these coefficients with machine accuracy. The coefficients remain constant and nonzero during the integration in the Z-direction, namely at each grid point $Z_j$. Increasing the number of collocation points has no influence on the results, i.e. the additional spectral coefficients are zero with machine accuracy.

*(ii) Oldroyd-B flow ($De > 0$) in a straight channel ($\Lambda = 1$).*

In this case, the analytical solution is given by Eqs. (3.16) and (3.17) which upon application of (B.1) gives zero spectral coefficients except for $\hat{\Psi}_1 = 9/16$, $\hat{\Psi}_3 = -1/16$, and $\hat{\tau}_{ns,1} = -3/2$.

Running the pseudospectral code for any value of *De* and $\eta$, using five collocation points (*M*=4) gives these coefficients with machine accuracy, at any axial distance $Z_j$. Increasing *M* generates zero additional spectral coefficients for all variables.

*(iii) Oldroyd-B flow ($De > 0$) in a hyperbolic channel ($\Lambda > 1$) for the uncoupled case ($\eta$=0) ignoring the Z-derivatives*

In this case, the underlying terms in Eqs. (3.6)-(3.9) are ignored, and there is no interaction between the momentum balance and the polymer extra-stresses. Consequently, the solution for the streamfunction and the pressure gradient is given by Eq. (3.16), while the solution for the polymer extra-stresses is simply found by substituting $\tilde{\Psi}(Y) = \Psi_S(Y) = 3Y/4 - Y^3/4$ in Eqs. (5.4a-c):

$$\tilde{\tau}_{ss}(Y) \approx \frac{9DeY^2/2}{\left(1+\frac{3De_m}{2}(1-Y^2)\right)^3}, \quad \tilde{\tau}_{ns}(Y) \approx \frac{-3Y/2}{\left(1+\frac{3De_m}{2}(1-Y^2)\right)^2}, \quad \tilde{\tau}_{nn}(Y) \approx -\frac{3(\Lambda-1)(1-Y^2)/2}{1+\frac{3De_m}{2}(1-Y^2)} \quad (B.1)$$

Similarly, Eqs. (5.5a-c) give:

$$\tilde{c}_{ss}(Y) \approx \frac{9De^2Y^2/2}{\left(1+\frac{3De_m}{2}(1-Y^2)\right)^3}, \quad \tilde{c}_{ns}(Y) \approx \frac{-3DeY/2}{\left(1+\frac{3De_m}{2}(1-Y^2)\right)^2}, \quad \tilde{c}_{nn}(Y) \approx \frac{1}{1+\frac{3De_m}{2}(1-Y^2)} \quad (B.2)$$

The functional form of the components of the polymer extra-stress and conformation tensors permits the analytical evaluation of the corresponding spectral coefficients, where the integrations are performed with the aid of *MATHEMATICA* software. For instance, the thirst three coefficients for $\tilde{\tau}_{ss}(Y)$ are:

$$\hat{\tau}_{ss,0} = \frac{9De(8+9De_m)}{8\sqrt{2}\,\Phi^{3/2}}$$

$$\hat{\tau}_{ss,2} = \frac{9De(4+9De_m)}{4\sqrt{2}\,\Phi^{3/2}} \quad (B.3)$$

$$\hat{\tau}_{ss,4} = \frac{-64}{3De_m^2(\Lambda-1)} + \frac{\sqrt{2}\left(1024+9De_m\left(256+3De_m\left(32-De_m(8-9De_m)\right)\right)\right)}{24De_m^2\Phi^{3/2}(\Lambda-1)}$$

where $\Phi \equiv 2+3De_m$ has been used for brevity. Multiplying Eq. (B.3) with *De*, gives the corresponding coefficient for $\tilde{c}_{ss}(Y)$. Running the code for $\eta$=0, $\Lambda$=6, and *De*=0.4 (thus, $De_m = (\Lambda-1)De = 2$) using 21 collocation points (M=20), the spectral coefficients of the components of the extra-stress and conformation tensors are calculated numerically. As in the previous two cases, the coefficients remain unchanged for any $Z_j$. The magnitude (absolute value) of the spectral coefficients for $\tilde{c}_{ss}$, $\tilde{c}_{ns}$ and $\tilde{c}_{ss}$ is shown in Figure A1, along with the corresponding analytically derived coefficients. Notice that only the non-zero coefficients are presented: the odd coefficients for $\tilde{c}_{ns}$ and the even coefficients for $\tilde{c}_{ss}$ and $\tilde{c}_{nn}$. The comparison, performed on a log-linear plot, clearly demonstrates that the code accurately reproduces the exact solution for all components of the conformation tensor. The linear decrease in the logarithm of the magnitude of the coefficients is also evident, though the small slope highlights the significant challenge of numerically resolving the polymer extra-stress or conformation components with high accuracy. Interestingly, even two spectral coefficients of the streamfunction generate an extra-stress tensor (or conformation tensor) that is considerably more demanding to resolve numerically.

These findings raise concerns about the capability of low-accuracy numerical schemes, such as first- or second-order finite difference methods, to accurately capture the polymer extra-stress or conformation tensor profiles in the vertical coordinate—particularly near the walls, where very steep gradients develop.

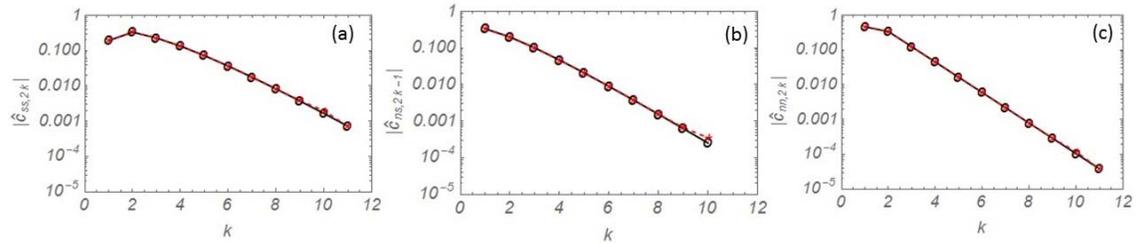

**Figure A1:** The magnitude of the non-zero spectral coefficients in log-linear scale for (a) $\tilde{c}_{ss}$; (b) $\tilde{c}_{ns}$; (c) $\tilde{c}_{nn}$; parameters are $\eta=0$, De=0.4, and $\Lambda=6$
Solid (black) lines: the analytically calculated coefficients (see Eq. (B.3)).
Dashed (red) lines: pseudospectral code results with 21 collocation points and ignoring the Z-derivatives for all variables.
The excellent agreement between the results demonstrates the spectral accuracy of the code.

*(iv) Oldroyd-B flow ( $De > 0$ ) in a hyperbolic channel ( $\Lambda > 1$ ) for the uncoupled case ($\eta$=0).*
This case is the same with case (iii), however, all the terms appearing into Eqs. (3.7)-(3.9) are taken into account. Specifically, for $\eta = 0$, $\Lambda = 6$ and $De = 0.4$ the simulations are performed using 21 collocation points (M=20). In Figure A2a-c, we present the magnitude of the spectral coefficients for $c_{ss}$, $c_{ns}$ and $c_{nn}$ close to the inlet ($Z = 0.1$), the middle of the channel ($Z = 0.5$), and the outlet ($Z=1$). This case also allows for the exact analytical solution for $c_{nn}(0,Z)$ (see

Eq. (3.26c)). Comparison of the analytical solution with the numerical solution shows agreement in seven significant digits.

Based on the simulations performed for the four cases described in this Appendix using the new set of lubrication equations, the spectral accuracy of our new pseudospectral code have been clearly demonstrated.

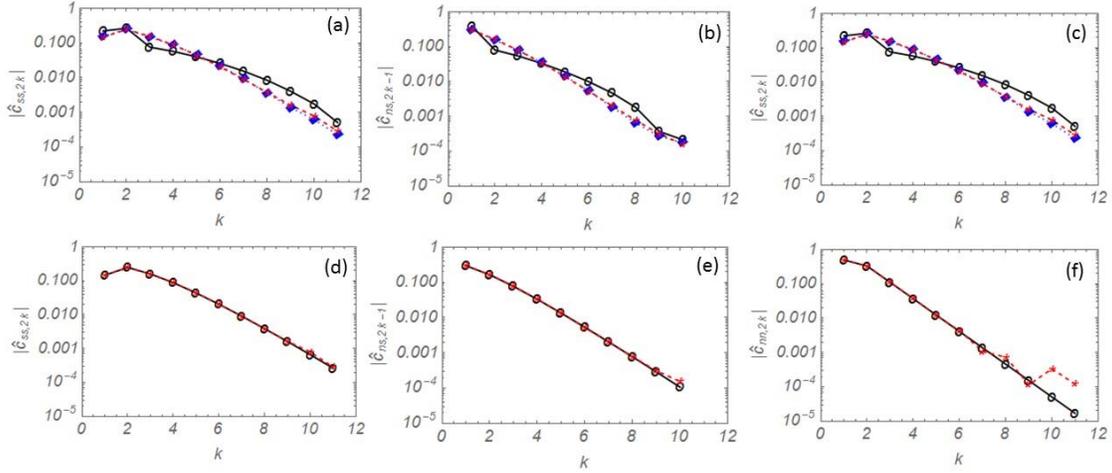

**Figure A2:** The magnitude of the spectral coefficients for (a) $\tilde{c}_{ss}$; (b) $\tilde{c}_{ns}$; (c) $\tilde{c}_{nn,}$ in log-linear scale; parameters are $\eta=0$, $De=0.4$, and $\Lambda=6$. The pseudospectral simulations are performed using 21 collocation points Black lines with open symbols: $Z=0.1$; Blue dotted lines with filled diamonds : $Z=0.5$; Red dashed lines with stars: $Z=1$. In Figures (d), (e), and (f) the spectral coefficients of the exact solution (solid black lines) are shown together with the corresponding coefficients at the exit of the channel ($Z=1$) (Red dashed lines with stars).

## Appendix C: The higher-order terms appearing in Eq. (6.1)

$$\pi_5(\eta) = -\frac{33048}{1001} + \frac{677484}{25025}\eta - \frac{2673}{175}\eta^2 + \frac{15552}{5005}\eta^3$$

$$\pi_6(\eta) = \frac{7776}{143} - \frac{15152832}{175175}\eta + \frac{7755912}{94325}\eta^2 - \frac{487296}{13475}\eta^3 + \frac{31104}{5005}\eta^4$$

$$\pi_7(\eta) = -\frac{1073088}{12155} + \frac{28588464}{119119}\eta - \frac{248460696}{728875}\eta^2 + \frac{5064514632}{20845825}\eta^3 - \frac{262737432}{2977975}\eta^4 + \frac{1119744}{85085}\eta^5$$

$$\pi_8(\eta) = \frac{2519424}{17765} - \frac{1536988608}{2540395}\eta + \frac{747265404096}{622396775}\eta^2 - \frac{187804006374573}{152487209875}\eta^3 +$$

$$\frac{617754255912}{871355485}\eta^4 - \frac{137154036672}{622396775}\eta^5 + \frac{6718464}{230945}\eta^6$$

57